\documentclass[prb,aps,amsmath,amssymb,superscriptaddress,longbibliography,notitlepage,twocolumn]{revtex4-2}
\usepackage{graphicx, multirow, outlines,ulem}
\usepackage{color}
\usepackage{xcolor}
\usepackage{subfig}
\usepackage{overpic} 
\usepackage{graphicx}
\usepackage{mathptmx}
\usepackage{hhline}
 \usepackage{physics}
\usepackage{hyperref}
\usepackage[mathscr]{eucal}

\usepackage{tikz}
\usepackage{xcolor}
\newcommand{\bo}{\begin{outline}}
\newcommand{\eo}{\end{outline}}

\newcommand{\ep}{\epsilon}

\newcommand{\qed}{\nobreak \ifvmode \relax \else
      \ifdim\lastskip<1.5em \hskip-\lastskip
      \hskip1.5em plus0em minus0.5em \fi \nobreak
      \vrule height0.75em width0.5em depth0.25em\fi}

\renewcommand{\thesection}{\arabic{section}}
\renewcommand{\thesubsection}{\thesection.\arabic{subsection}}
\renewcommand{\thesubsubsection}{\thesubsection.\arabic{subsubsection}}

\usepackage{amsmath}
\usepackage{amssymb}
\usepackage{graphicx}
\usepackage{xcolor}

\definecolor{iopred}{RGB}{170,0,0}



\begin{document}
\title{Topological properties of curved spacetime extended Su–Schrieffer–Heeger model }
\author{Priyanuj Rajbongshi}

\email{ph24d004@iittp.ac.in}
\author{Ranjan Modak}
\email{ranjan@iittp.ac.in}
\affiliation{Department of Physics, Indian Institute of Technology Tirupati, Tirupati 517619, India}

\begin{abstract}

      The Su-Schrieffer-Heeger (SSH) model, a prime example of a one-dimensional topologically nontrivial insulator, has been extensively studied in flat space-time. In recent times, many studies have been conducted to understand the properties of the low-dimensional quantum matter in curved spacetime, which can mimic the gravitational event horizon and black hole physics. However, the impact of curved spacetime on the topological properties of such systems remains unexplored. In this work, we investigate the curved spacetime (CST) version of the extended SSH model,  
      by introducing a position-dependent hopping parameter. 
      The extended SSH model already exhibits topological phases and the associated phase transitions. Different topological markers suggest that for the same choice of parameters, the CST version of the model retains the imprint of the same topological phases and transitions. Furthermore, the topologically non-trivial phase of the CST model hosts zero-energy edge modes, which are spatially asymmetric in contrast to those of the conventional SSH model.
      We find that at the topological transition points between phases with different winding numbers, a critical slowdown takes place for zero-energy wave packets near the boundary, indicating the
presence of a horizon, and interestingly, if one moves even a slight distance away from the topological transition points, wave packets start bouncing back and reverse direction before reaching the boundary. 
Moreover, we have also quantified the time scale of the critical slowdown of the wavepacket across different winding-number transition phases. A semiclassical description of the wave packet trajectories also supports these results.

\end{abstract} 
\maketitle

\section{Introduction}
One of the most astonishing predictions of Einstein’s theory of general relativity~\cite{einstein2005grundlage} is the potential existence of black holes, i.e., space-time regions from where nothing can escape. From then on, there have been efforts to simulate a black hole horizon and the relevant curve spacetime (CST) physics in the laboratories. In 1981, Unruh proposed a sonic horizon, which is based on the observation that sound waves in flowing fluids, in appropriate conditions, can be described by the same wave equation as a scalar field in a curved space-time \cite{unruh1981experimental}. The acoustic horizon~\cite{PhysRevLett.114.036402} occurs if the velocity of the fluid exceeds the speed of sound within the liquid, acting on sound waves exactly as a black hole horizon does~\cite{ PhysRevD.66.044019}.  Further,  there exist proposals for black hole analogs~\cite{reznik2000origin, schutzhold2002dielectric}, sonic black-hole analog in Bose-Einstein condensates~\cite{PhysRevLett.105.240401,PhysRevLett.85.4643}, analog Hawking radiation~\cite{Tian_2022}, Gibbons-Hawking effect~\cite{ PhysRevLett.91.240407} and other curved spacetime phenomena analogs based on light propagation in dielectric media~\cite{ novello2000geometrical, novello2001effective}, liquid Helium 3 \cite{volovik2001superfluid}, Bose-Einstein
condensates ~\cite{PhysRevA.70.063615, PhysRevD.105.124066}, classical electronics circuits~\cite{kollar2019hyperbolic}, floquet quantum systems~\cite{lapierre2024floquetengineeredinhomogeneousquantum,PhysRevResearch.2.023085}.
These works have provided a setup for studying the CST physics in laboratories, which has deepened the understanding of the curved spacetime and  gravity, e.g., one of the milestones of these studies was to reveal the relation between the Sachdev-Ye-Kitaev model \cite{PhysRevLett.70.3339} and (1 + 1)D Jackiw-Teitelboim
gravity  ~\cite{PhysRevD.94.106002,kitaev2018soft}. Only in the last few years, there have been some studies that have focused on the condensed matter properties in CST lattice systems, asking extremely fundamental questions like what will happen to the fate of thermalization \cite{mertens2022thermalization} and localization~\cite{li2023anderson} in CST lattice models~\cite{morice_2022,morice2021synthetic}.

On the other hand, one of the main goals of condensed matter physics is to deal with different 
phases of matter. Traditionally, phase transitions were characterized by order parameters within Landau’s free-energy theory framework and symmetry breaking. 
In the past few decades, the identification of new phases of
matter that did not break any symmetry, nor could be characterized by the usual
order parameters, has led to the appearance of topology in condensed matter systems.
While 2D electronic systems, which displayed a quantized Hall conductance~\cite{PhysRevLett.49.405}, are probably 1st known example of such a condensed matter system where topology plays a crucial role, in recent days,  a new class
of electronic materials such as topological insulators~\cite{ti1,ti3},
topological crystalline insulators~\cite{tc1,tc2}, and topological semimetals~\cite{sm1,sm3} have emerged as a material having nontrivial Bulk band topology that can be exploited for application in low-power consumption electronic and spintronic devices due to the robustness of their edge states to defects, which are topologically protected.  In this context, to understand how these materials behave in
realistic situations, it is important to first understand simple toy models that show such a topological phase transition. One such example of a toy model is the one-dimensional (1D) Su-Schrieffer-Heeger (SSH) model, which is an example of a topological insulator~\cite{su1980soliton}. 
It is a tight-binding model of non-interacting spinless electrons confined in a dimer chain, and has been extensively studied both theoretically~\cite{PhysRevB.21.2388, PhysRevB.26.4278} and experimentally~\cite{meier2016observation} in the recent past~\cite{RevModPhys.60.781,joshi2025adiabatic}. The SSH model was initially introduced to describe a 1D  chain of polyacetylene, on which electrons hop with staggered hopping amplitudes. There are two sites in each unit cell of this model; thus, it is a two-band model, and has winding number as its corresponding topological invariant. 
In particular, the number of edge states on a boundary of the system has a one-to-one correspondence to the winding number associated with this model. Recent studies have shown that extended versions of the SSH model can realize winding numbers beyond those accessible in the standard SSH model through the use of enlarged unit cells, staggered long-range hoppings, or further-neighbor couplings \cite{Maffei_2018, Malakar_2023}. These higher-winding number phases have also been implemented in acoustic metamaterial \cite{PhysRevApplied.19.054028}, circuit heterostructures \cite{Guo_10.1063/5.0283673}, and sonic semimetal \cite{PRL_high_winding_2025} platforms, where they give rise to multiple edge states and momentum-dependent bulk-edge correspondence.

In this work, our main goal is to connect these two extremely different branches of physics, 1) 
curved spacetime and black hole, and 2) topological systems in condensed matter. We propose a CST lattice model inspired by the two-band extended SSH models. First, we investigate whether we can see the event horizon-like dynamics or not. Next, we investigate whether the CST version of the extended SSH models has a topologically non-trivial phase or not. Remarkably, we answer both questions affirmatively in this manuscript. Using different topological markers we show that the CST-SSH model and its extended version also show a topological phase transition. Most interestingly,  exactly at the transition point, a critical slowdown takes place for zero-energy wave packets, which indicates the
presence of a horizon. It is also important to clarify that, in contrast to recent studies of curved-space photonic lattices, where genuine geometric curvature can induce topological phases, edge states, and localization effects \cite{PhysRevA.96.041804}, the CST lattice considered here does not require physical spatial curvature. Instead, an effective curved spacetime emerges from position-dependent hopping amplitudes. In the low-energy continuum limit, this system maps onto a Dirac equation in curved spacetime.

This paper is organized as follows: We define our CST-SSH models in Section~\ref{model}. Section~\ref{wp_sc} shows the wave packet dynamics and semi-classical analysis of the dynamics.We investigate the gap of the spectrum and also the zero-energy states in Sec.~\ref{gap}. We dedicate the Sec.~\ref{signature} for different topological markers to check the topological nature of the models. Section.~\ref{quench} shows the effect of quench across different topological phases in CST-SSH. We then quantify the slowdown parameter in Sec.~\ref{slowdown} for the extended CST-SSH.
Finally, we summarize our findings in Sec.~\ref{summary}.

\section{Curved spacetime SSH models\label{model}}
In some recent studies~\cite{morice_2022,morice2021synthetic}, an equivalence between the Dirac equation in curved spacetime and a tight-binding condensed matter system with power law position-dependent hopping has been demonstrated. 
As suggested, we can establish a direct connection between the continuum field theory and a condensed matter system by considering a simple tight-binding Hamiltonian in real space with nearest-neighbor (nn) position-dependent hopping:
\begin{equation}
    \mathcal{H}=  -\sum_{n=1}^{N-1}t_n\big(  c^\dagger_{n} c_{n+1}+h.c.\big),
    \label{eq: CST TB equation}
\end{equation}
where $c^\dagger_{n}$ and $c_n$ are fermionic-creation and annihilation operators, and $t_n$ is a position-dependent hopping parameter. If $t_n=t$ (the nn hopping amplitude is constant for all sites), then the Hamiltonian can be diagonalized very easily in the momentum space $k$. The Hamiltonian will read as, $\mathcal{H}=\sum_k\epsilon(k) c^{\dag}_kc_k$, where $c^\dagger_{k}$ and $c_k$ are creation and annihilation operators in momentum space, and $\epsilon(k)=-2t\cos k$. In case of position-dependent hopping $t_n (~\approx t\big(\frac{n
}{N}\big)^\sigma)$, and in the limit $ N\rightarrow\infty$, 
the difference between the two adjacent hopping amplitudes, 
$t_{n+1}-t_n= t\Big[(x+\frac{1}{N})^\sigma-(x)^\sigma\Big]$, 
where, $x=\frac{n}{N}$. Now, the 1st term, in the limit $Nx$ large, read as $\Big(x+\frac{1}{N}\Big)^\sigma=  x^\sigma\Big(1+\frac{\sigma}{Nx}+ \mathcal O((\frac{1}{Nx})^2)\Big)
\text{. Therefore,  }(t_{n+1}-t_n)\approx t\Big(\frac{x^{\sigma-1}\sigma}{N}\Big)$
. Thus, in the thermodynamic limit, $N \xrightarrow{}\infty$ and for the bulk, i.e., $n\gg 0$, the hopping variation in neighboring sites is effectively negligible, i.e. for a short segment of the lattice, the hopping is effectively constant. Thus, 
we can approximate the local band structure for the bulk  as $\epsilon(n,k) \sim -2t_n\cos k$, and the local Fermi velocity as $v_F(n)\sim \pm 2t_n$. Thus, the local velocity varies with spatial coordinates, as the hopping here is position-dependent. In the continuum limit, these kinds of lattice models are equivalent to a massless Dirac field  with the two-component spinor $\hat \Psi= (\hat \psi_+,\hat \psi_-)^T$ that obeys the following time-evolution equation~\cite{morice_2022},
\begin{equation}
  \partial_\tau\hat\Psi=\Sigma_3\Big(v(x)\partial_x+\frac{1}{2}\frac{dv}{dx}\Big)\hat\Psi,  
  \label{eq:dirac eq in cst}
\end{equation}
where, $\Sigma_i,(i=1,2,3)$ are the Pauli matrices and $v(x)$ is a position-dependent velocity, relating to the hopping as $v(x)=2t(x)$. It is identical to the Dirac equation on  (1+1) D spacetime in the massless limit. Thus, Eq.~\ref{eq: CST TB equation} is the low-energy discrete version of a Dirac Hamiltonian in the continuum limit with the equation of motion ~\ref{eq:dirac eq in cst}, and with the Rindler metric 
\begin{equation}
    ds^2=-v^2(x)d\tau^2+dx^2.
    \label{eq: rindler metric}
\end{equation}
At $v(x)=0$, i.e., $t(x)=0$, this Rindler metric possesses an event horizon, as at that point, the local speed of light goes to zero. If we let $t(x)\propto x^\sigma$, where $\sigma$ is indicating the warping of spacetime \cite{li2023anderson}, equivalently, in the lattice model ~\eqref{eq: CST TB equation} $t_n=(\frac{n}{N-1})^\sigma$, at $n/N\to 0$ limit,  the lattice model will possess event horizon like dynamics for $\sigma \geq 1$: an gaussian wave-packet propagating in this lattice will suffer eternal slowdown and never reaches the origin, thus effectively forming an event horizon like critical slowdown at the origin in the lattice model.
\\In a similar spirit, in this work, we investigate whether we observe the horizon-like physics in a two-band topological system with next nearest neighbor hopping, and what would happen to the topological characteristics after introducing position dependent hopping. Hence, we study an extended version of the SSH model, where the hopping is not just restricted to between nearest neighbors, but higher-order hopping has also been allowed. In the context of the standard SSH model, a similar model was studied in Ref.~\cite{PhysRevB.110.054312,jin2024topologicalfinitesizeeffect}, known as the extended SSH model. Here we study the CST version of such models, where hopping is position-dependent, and the model reads as,\\
\begin{equation}
\begin{aligned}
H &= \sum_{n=1}^{N}t_1(n)\,c^\dagger_{n,A}c_{n,B}
  \;+\;\sum_{n=1}^{N-1}t_2(n)\,c^\dagger_{n,B}c_{n+1,A}\\
  &\quad+t_3(n)\,c^\dagger_{n,A}c_{n+1,B}
  \;+\;\sum_{n=1}^{N-2}t_4(n)\,c^\dagger_{n,B}c_{n+2,A} + h.c.,
\end{aligned}
\label{eq:ext_CST_SSH}
\end{equation}
where $c^\dagger_{n,A}$, $c_{n,A}$ ($c^\dagger_{n,B}$, $c_{n,B}$) are fermionic creation and annihilation operators at the A sub-lattice (B sub-lattice) of the nth unit cell, and, $t_1(n)= t_1\big(\frac{2n-1}{2N-1}\big)^\sigma$,  $t_2(n)= t_2\big(\frac{2n}{2N-1}\big)^\sigma$ are the nearest neighbor intracellular and intercellular position-dependent hoppings, and, $t_3(n)=t_3\big(\frac{2n-1}{2N-1}\big)^\sigma$,  $t_4(n)= t_4\big(\frac{2n}{2N-1}\big)^\sigma$. For $\sigma=0$, the Hamiltonian will have translational invariance. Hence, it can be Fourier transformed to momentum $k$ space, where the winding number transition can happen when the components of the Bloch vector become zero, i.e., $d_x(k)=t_1+(t_2+t_3)\cos k+t_4\cos2k$ and, $d_y(k)=(t_2-t_3)\sin k+t_4\sin2k$ both becomes 0, which gives us conditions: 
\begin{equation}
\begin{aligned}
t_1 + t_2 + t_3 + t_4 &= 0,\\
t_1 - (t_2 + t_3) + t_4 &= 0,\\
t_4(t_1 - t_4) + t_3(t_3 - t_2) &= 0.
\end{aligned}
\label{eq:constraints}
\end{equation}

at gap closing momentum $k_c=0,~ \pi~ $ and for $k_c$ satisfying $\cos k_c=\frac{t_3-t_2}{2t_4};$ ($t_4\neq 0, |\frac{t_3-t_2}{2t_4}|\leq 1)$, respectively, where we can observe winding number transitions. 
To gain insight into the topological properties , we also consider a limiting case of the Hamiltonian \ref{eq:ext_CST_SSH}, with $t_3=t_4=0$, which can be considered as the curved-space-time version of the standard SSH model.


\section{Wave-packet vs semiclassical dynamics\label{wp_sc}}
\begin{figure*}[!tb]
  \centering
  
  \subfloat[]{%
    \includegraphics[width=0.34\textwidth]{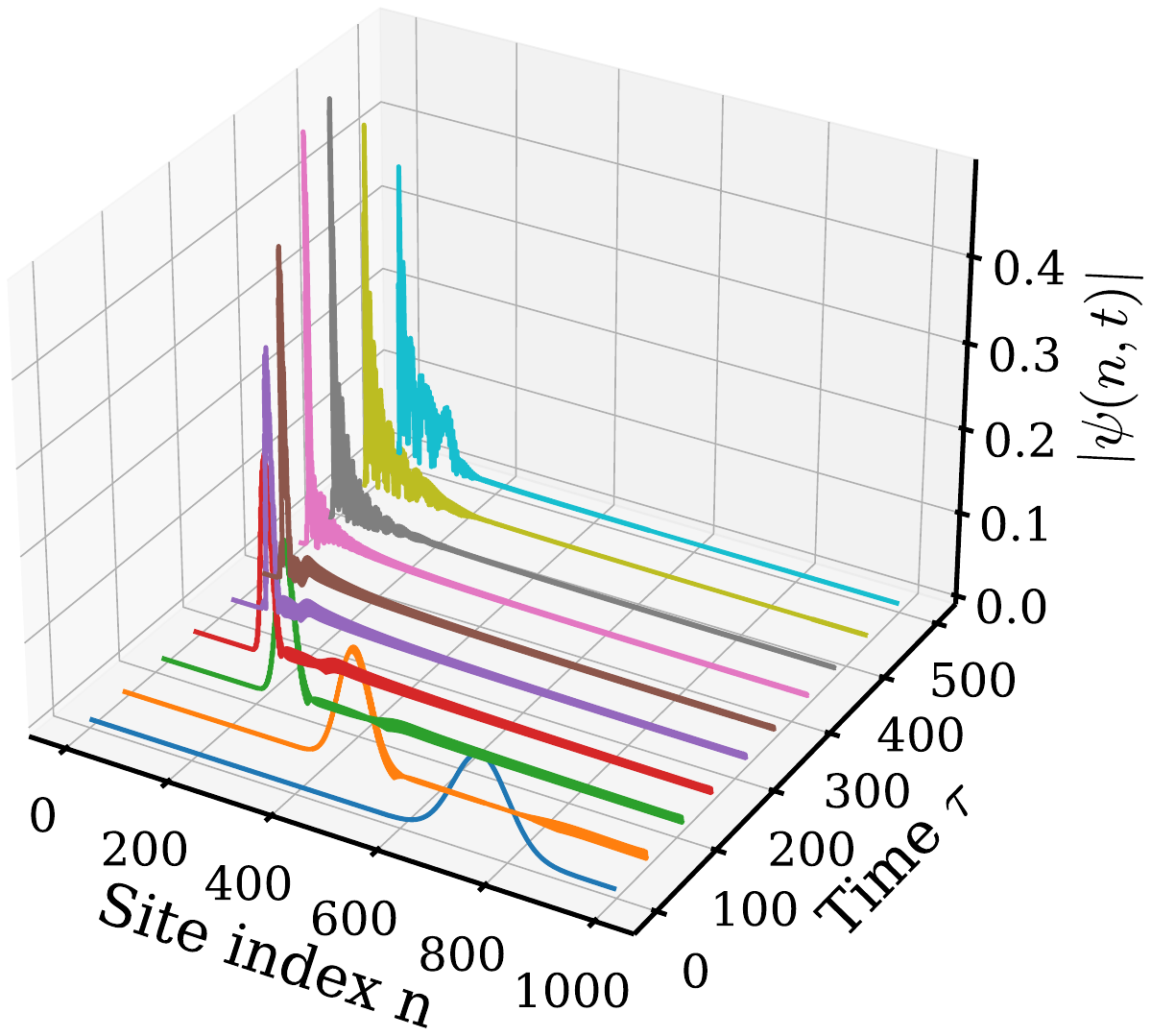}%
     \label{fig:ext_gau_pmmp}}%
  \subfloat[]{%
  \includegraphics[width=0.34\textwidth]{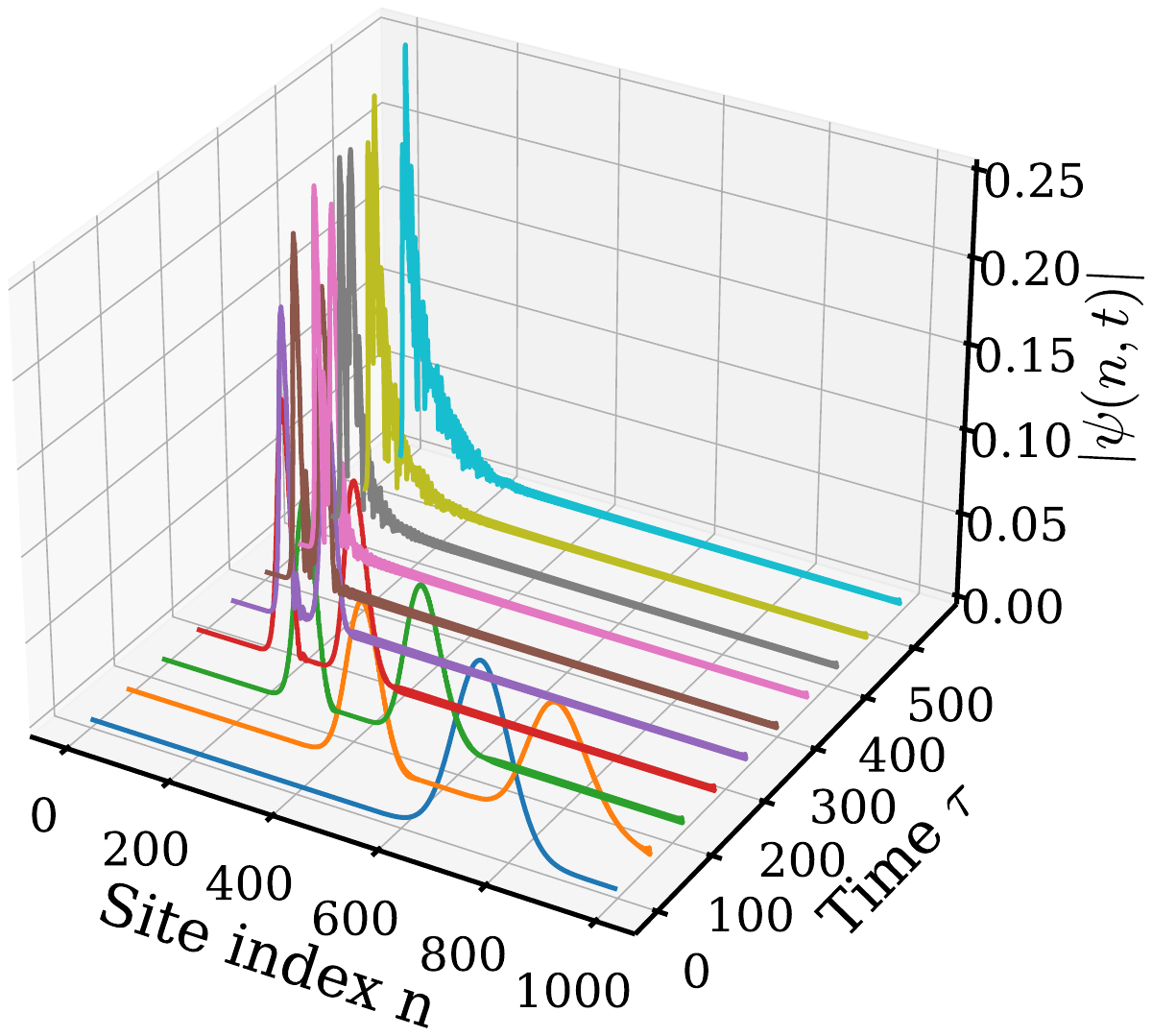}%
    \label{fig:ext_gau_pppp}}%
  \subfloat[]{%
    \includegraphics[width=0.34\textwidth]{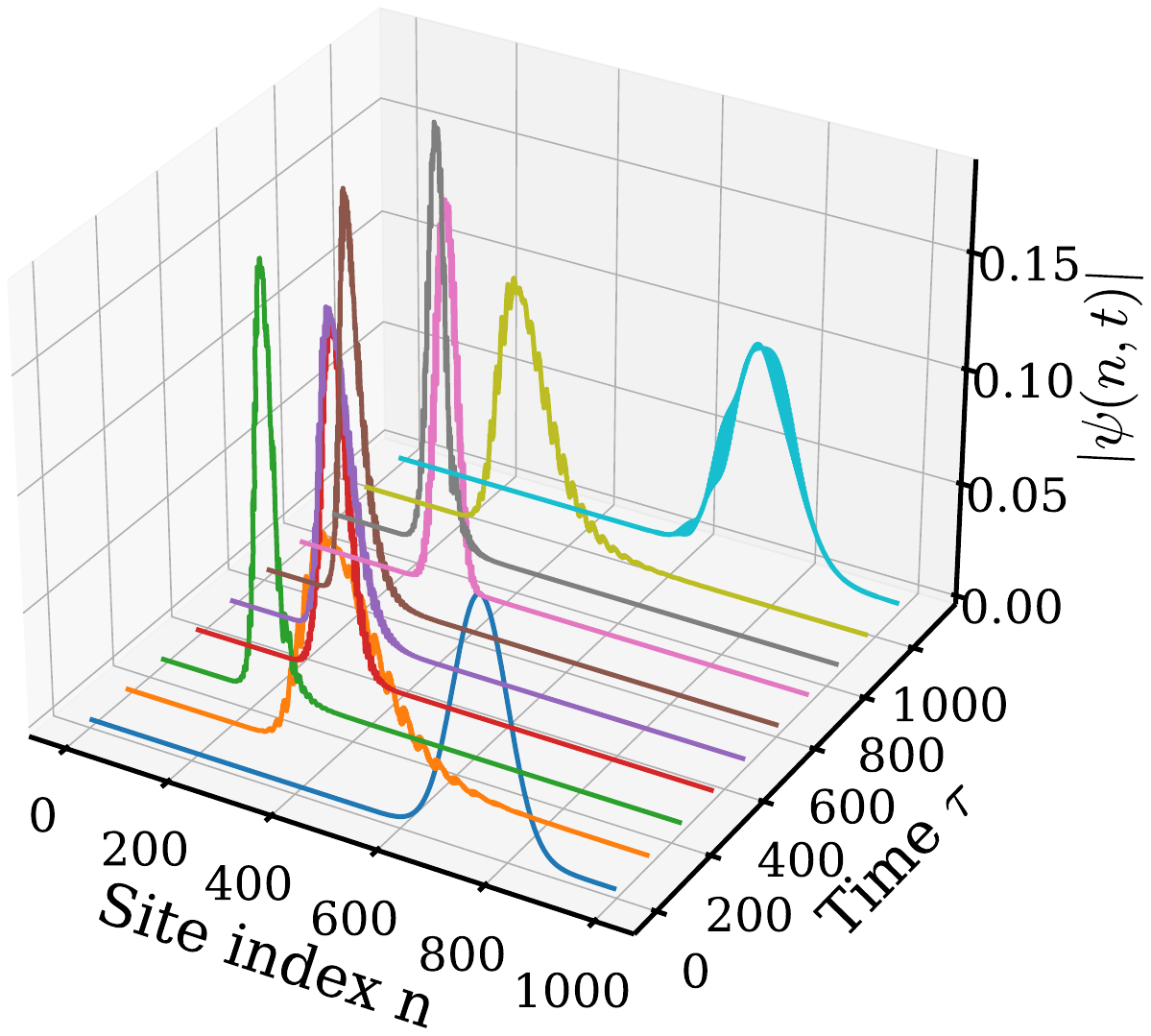}%
    \label{fig:ext_gau_winding2}}\\[-1.75cm]
  \subfloat[]{%
    \includegraphics[width=0.35\textwidth]{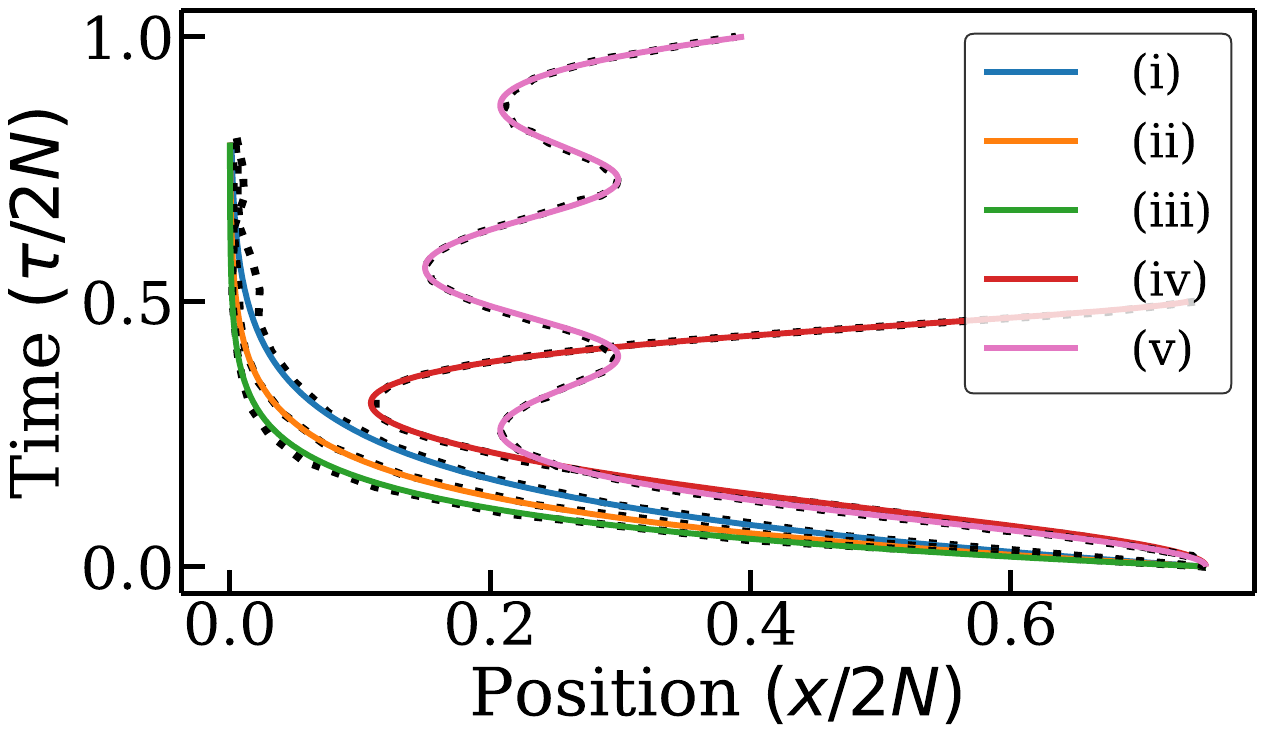}%
    \label{fig:ext_sc_vs_num}}
  \subfloat[]{%
      \includegraphics[width=0.34\textwidth]{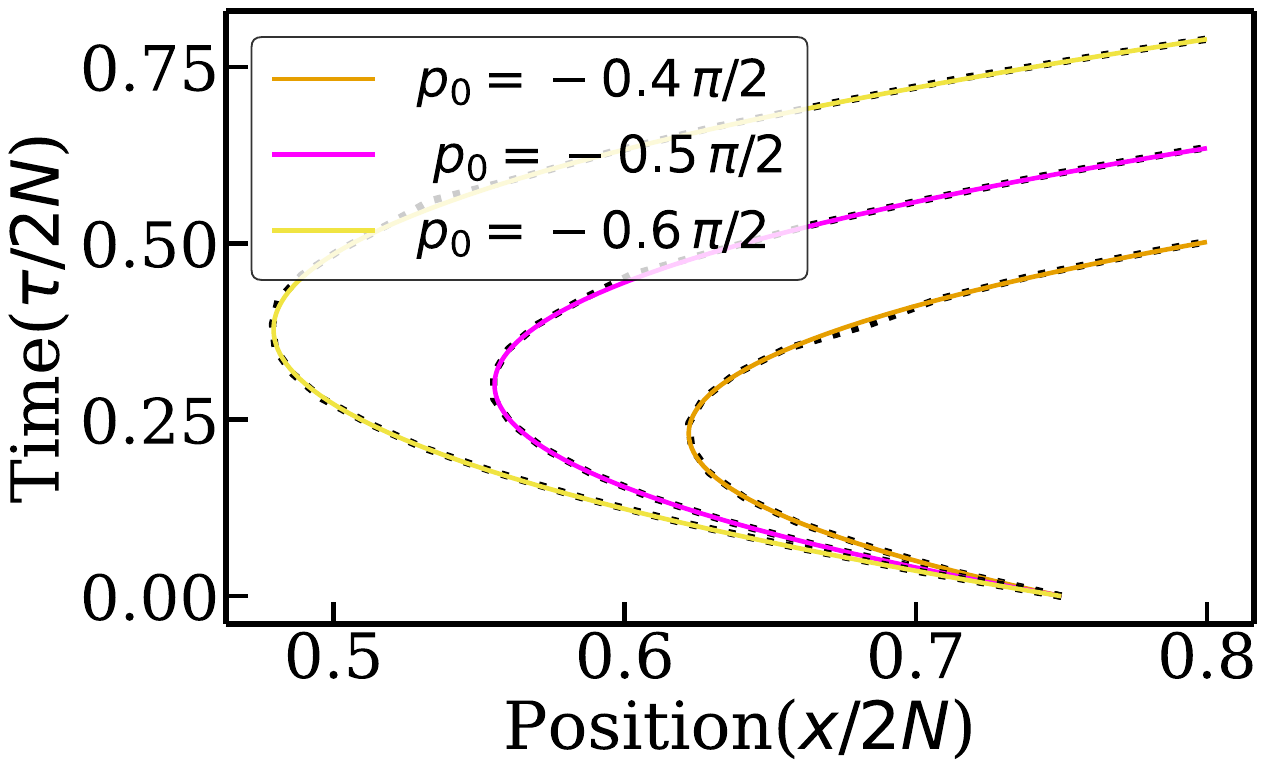}%
      \label{fig:sc vs num}}%
    \subfloat[]{%
      \includegraphics[width=0.32\textwidth]{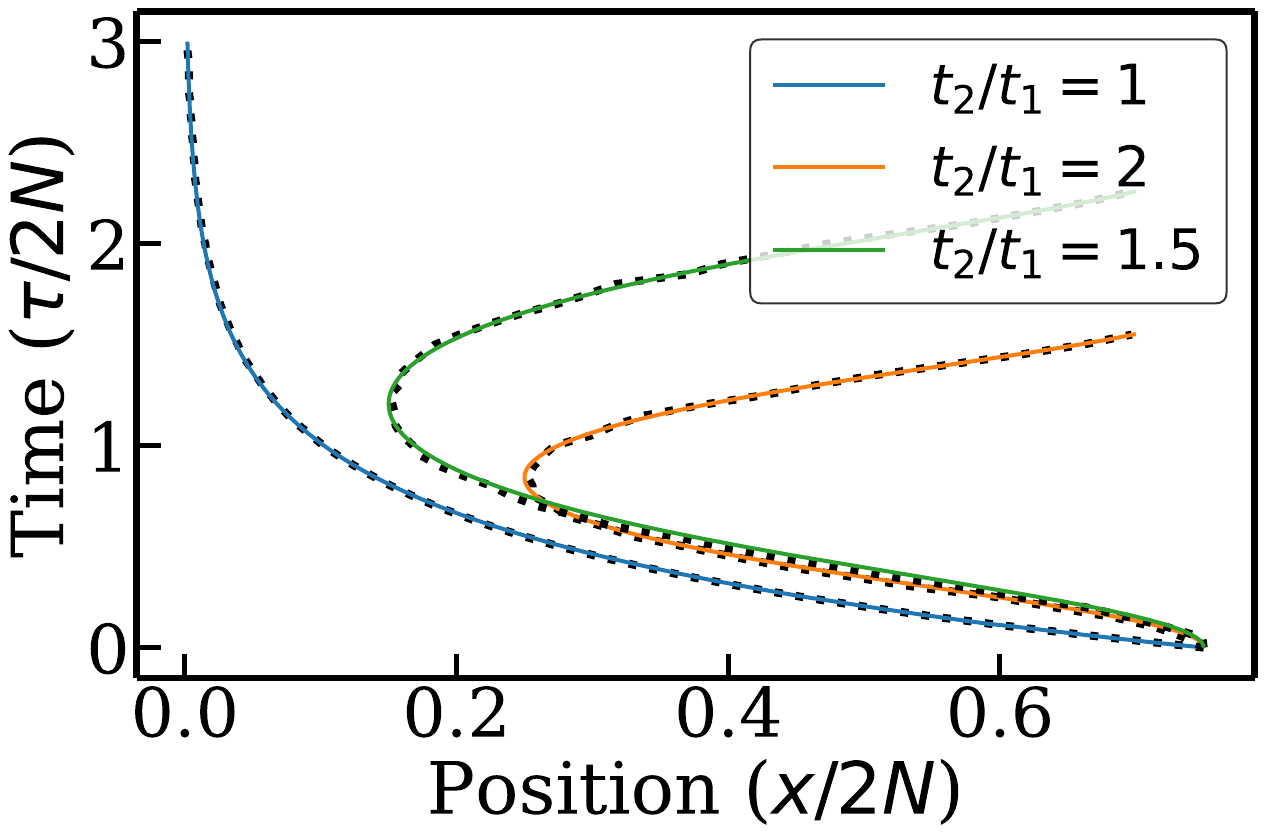}%
      \label{fig:sc vs num ratio}}
    \\[-4cm]
  \caption{Time evolution of Gaussian wave packet in the Extended CST-SSH Hamiltonian of system size $2N=1000$ (500 unit cells) with $\sigma=1, \omega=50, x_0=750$: (a) obeying the relation $t_1 - (t_2 + t_3) + t_4 = 0$ with hopping $(t_1,t_2,t_3,t_4) =(3,-2,1,-4) $, with $|p_0|=\pi/2$, (b) obeying the relation $t_1 +t_2 + t_3 + t_4 = 0$ with hopping $(t_1,t_2,t_3,t_4) = (0,1,1,-2)$, at $|p_0|=\pi$ (c) extended CST-SSH having winding no 2, with hopping parameters $(t_1,t_2,t_3,t_4) = (0,1,1,-3)$ at $|p_0|=\pi$ (d) peak position vs. time for different combinations of hopping and initial momentum$ (t_1,t_2,t_3,t_4,p_0)$ : (i) $(0,1,1,-2,\pi)$, (ii) $(3,-2,1,-4,\pi/2)$, (iii) $(3,1,1,3,\pi/4)$ (iv) $(3,-2,1,-3,\pi/2)$, and, (v) $(0,1,1,-3,\pi)$. (i), (ii) and (iii) suffer eternal slowdown; whereas, (iv) and (v) return before reaching the origin. For $\sigma=1, \omega=25, x_0=750$: (e) peak position of the Gaussian wave packet vs.\ time for different $p_0$, at $\sigma=1$, having $(t_1,t_2,t_3,t_4)=(1,1.9,0,0)$; (f) peak position vs.\ time for different $t_1/t_2$ ratios with $(t_3,t_4)=(0,0)$ at $p_0=-\frac{\pi}{2}$. Dotted lines are numerical; solid lines are semiclassical predictions.}
  \label{fig:MCD}
\end{figure*}

The first question we try to address here is whether we see any event {horizon-like} physics for the Hamiltonian Eq.~\eqref{eq:ext_CST_SSH}. Hence, we first study the wave-packet dynamics. 
In this section,  we study the wave packet dynamics under the CST-SSH Hamiltonian of a Gaussian wave packet and compare the results with the semiclassical set of coupled differential equations that govern wave packet trajectories. We consider the following Gaussian wave packet,
\begin{equation}
    \psi(x,\tau=0)=\frac{1}{\sqrt[4]{\pi\omega^2}}e^{-\frac{1}{2}\big(\frac{x-x_0}{\omega}\big)^2}e^{ip_0x},
\end{equation}
 which describes a normalized gaussian wave packet at time $\tau=0$ centered at position $x_0$ of the lattice, and  having initial momentum of $p_0$, $\omega$ is the width of the wave packet. We observe that, for the parameter values satisfying any of the conditions at Eq. \ref{eq:constraints}, the wavepacket having the initial momentum of $p_0=k_c/2$ suffers a critical slowdown down while approaching towards the origin, and , does not return (Figs.\ref{fig:ext_gau_pmmp}, \ref{fig:ext_gau_pppp}), for $\sigma \geq1$. On the other hand, for any other combinations of $(t_1,t_2,t_3,t_4)$, it comes back before reaching the origin \ref{fig:ext_gau_winding2},  which we identify as a `turning point'. This $x_{\min}$ depends on the details, e.g., initial momentum $p_0$, the Hamiltonian parameters $\sigma$, $t_1,t_2,t_3$, and $t_4$.
 In case of the CST version of the standard SSH, we find that if $t_1\neq t_2$, the wave-packet first propagates towards an edge and then returns from some point $x_{\min}$, and, for $t_1=t_2$ and $p_0=-\pi/2$, it eternally slows down while evolving as it approaches the edge $x=0$, and never returns. We can identify it as the asymptotic localization of wave packets at the origin, which mimics the key feature expected for wave packet dynamics in the presence of an event horizon.

  \captionsetup[subfigure]{position=top,skip=0pt,aboveskip=0pt,belowskip=0pt}

\subsection{Semiclassical dynamics for the Extended CST-SSH model}
We can dive deeper into the time evolution of the wave packet by obtaining the semiclassical trajectories of the Hamiltonian of the extended version of the CST-SSH model. For a position-dependent extended SSH model, the two recursive energy eigenvalue equations can be written as,
\\
\begin{eqnarray}
      t_4(n-2)\psi_{B,n-2}+ t_2(n-1)\psi_{B,n-1}+ t_1(n)\psi_{B,n}&+&t_3(n)\psi_{B,n+1} \nonumber\\
      &=&\epsilon\psi_{A,n}
      \label{eq: Sublattice A semiclassicalextended}
\end{eqnarray}
  
 \begin{equation}
     t_3(n-1)\psi_{A,n-1}+t_1(n)\psi_{A,n}+ t_2(n)\psi_{A,n+1}+t_4(n)\psi_{A,n+2}=\epsilon\psi_{B,n}
      \label{eq: Sublattice B semiclassical extended}
 \end{equation}
 Following the reference \cite{morice_2022}, we introduce continuous function $\tilde \psi_{A/B}(x_n)$ which is related to the discrete $\psi_{A/B,n}$ of the lattice model as: $\tilde \psi_{A/B}(x_n)=\psi_{A/B,n} $, where,  $x_n=\frac{n}{N-1}$ is the position of the $n$-th unit cell in the conituum space, with  $\tilde \psi_A(x_n)-\tilde\psi_B(x_n)=\delta x=\frac{1}{2N-1} $, the minimum distance on the lattice. Expanding the wave functions in Taylor series, we get a mapping between the discrete $\psi_{A/B,n\pm 1}$ with continuous $\tilde \psi_{A/B}(x_n)$ as $\psi_{A/B,n\pm 1}=e^{\pm i2\delta x \hat{p}}\tilde \psi_{A/B}(x_n)$. The factor of two comes as changing the unit cell from $n$ to $n\pm 1$ corresponds to moving $2\delta x$ in space to the right or left, and the minimum length scale on the lattice is $\delta x$.
 We then end up with  the following pairs of equations,
 \begin{equation}
\begin{aligned}
\Bigl(
  &t_4(\hat{x}-4\delta x)e^{-4 i \delta x \hat{p}}
  + t_3(\hat{x})e^{2 i \delta x \hat{p}}
\\
  &+ t_2(\hat{x}-2\delta x)e^{-2 i \delta x \hat{p}}
  + t_1(\hat{x})
\Bigr)\tilde\psi_B(x,\tau)
= i\,\partial_\tau \tilde\psi_A(x,\tau)
\label{eq:continuum-psi-A}
\end{aligned}
\end{equation}

 and, 
\begin{equation}
\begin{aligned}
\Bigl(
  t_4(\hat{x})e^{4 i \delta x \hat{p}} 
  &+ t_3(\hat{x}-2\delta x)e^{-2 i \delta x \hat{p}}
\\
  &+ t_2(\hat{x})e^{2 i \delta x \hat{p}}
  + t_1(\hat{x})
\Bigr)\tilde\psi_A(x,\tau)
= i\,\partial_\tau \tilde\psi_B(x,\tau)
\label{eq:continuum-psi-B}
\end{aligned}
\end{equation}

 where, $\tilde \psi_{A/B}(x,\tau)=\tilde \psi_{A/B}(x)e^{-i\ep \tau}$. Combining the two equations, we get the continuum Hamiltonian as a $2\times 2$ matrix
 

 \[
\tilde H = \begin{pmatrix}
0 & T(\hat x,\hat p) \\[1mm]
T^*(\hat x,\hat p) & 0
\end{pmatrix},
\]

where, $T(\hat x,\hat p)=t_1(\hat x)+e^{-2i\delta x\hat p}t_2(\hat x)+t_3(\hat x)e^{2i\delta x\hat p}+e^{-4i\delta x\hat p}t_4(\hat x)$, 
which acts on the spinor $\Psi(x,\tau) = \bigl(\begin{smallmatrix} \tilde\psi_A(x,\tau)\\ \tilde\psi_B(x,\tau) \end{smallmatrix}\bigr)$. 
Now, neglecting the commutation relation between $\hat x$ and $\hat p$, and then, rescaling the momentum and time as $p\rightarrow p/\delta x$ and $\tau\rightarrow \tau/\delta x$, we could express the energy as follows, and with it, we can obtain the equation of motion.
The semiclassical equations of motion (in one dimension) read:
 \begin{equation}
\dot{x} \;=\; \frac{\partial E_{\pm}}{\partial p},
\qquad
\dot{p} \;=\; -\,\frac{\partial E_{\pm}}{\partial x}.
\label{eq: x dot p dot}
\end{equation}
Here, The energy expression, $$E_{\pm}(x,p)=\pm|T(x,p)|=\pm \sqrt{T_R^2(x,p)+T_I^2(x,p)},$$ where, $T_R=t_1(x)+(t_2(x)+t_3(x))\text{cos}(2p)+t_4(x)\text{cos}(4p)$ and $T_I(x,p)=(t_3(x)-t_2(x))\text{sin}(2p)-t_4(x)\text{sin}(4p)$ are the real and imaginary part of $T(x,p)$. Now, after substituting the values of $T_R$ and $T_I$, one can obtain the equations of motion as:
\begin{equation}
    \dot x= \pm\frac{1}{\sqrt{T_R^2(x,p)+T_I^2(x,p)}}(T_R\partial_p T_R+T_I\partial_p T_I)
\end{equation}
\begin{equation}
    \dot p= \mp\frac{1}{\sqrt{T_R^2(x,p)+T_I^2(x,p)}}(T_R\partial_x T_R+T_I\partial_x T_I),
\end{equation}
where, $\partial_p$ and $\partial_x$ are the partial derivatives with respect to $p$ and $x$, respectively. Here, the $\mp$ signs just indicate the direction of the propagation of the trajectories. Fig. \ref{fig:ext_sc_vs_num} shows the comparison between the numerical time evolution ( dotted black lines) and the semiclassical equations of motion ( colored lines ) to obtain the peak positions of the wave packets at various times for different hopping parameters. The semiclassical calculations show excellent agreement with the numerical time evolution of the Gaussian wavepackets.
Semiclassical trajectories can also allow us to predict the turning point. One can analytically obtain the turning point by taking the conserved nature of the average energy of wave packet from the semiclassical calculation.
One can obtain the analytical value of the turning point by equating the energy expression at $(x_0,p_0)$ and at $(x_{\min}, p)$, where choosing an appropriate $p$ makes $\dot x=0$. For $p=0$, the turning position reads as:
\begin{widetext}
\begin{equation}
\label{eq:xmin_extended}
x_{\min}
= x_0 \Biggl[
  \frac{(t_1 + (t_2+t_3)\cos 2p_0 + t_4\cos 4p_0)^2
        +((t_3-t_2)\sin 2p_0 - t_4\sin 4p_0)^2}
       {(t_1 + t_2 + t_3 + t_4)^2}
\Biggr]^{\frac{1}{2\sigma}}
\end{equation}
\end{widetext}


In Eq.~\eqref{eq:xmin_extended}, the expression in the numerator is exactly the squared band energy of the corresponding homogeneous extended SSH chain evaluated at momentum $p_0$, while the denominator is the squared band energy at $k=0$. It has been derived by equating the energy at $(x_0,p_0)$ and at $(x_{\min},p)$, and then choosing $p=0$ so that the semiclassical velocity $\dot{x} = \partial_p E(x,p)$ vanishes at the turning point. We have also assumed $( t_1 + t_2 + t_3 + t_4 ) \neq 0$, so that the denominator remains finite and the turning point is determined entirely by the value of the numerator. For fixed hopping amplitudes and fixed $x_0$, the turning position $x_{\min}$ is therefore controlled by how large the homogeneous band energy at the initial momentum $p_0$ is: when this energy decreases, the numerator of Eq.~\eqref{eq:xmin_extended} becomes small and the packet penetrates deeper; when the numerator vanishes, $x_{\min}$ becomes zero.

The condition that the numerator is zero is precisely the bulk gap closing condition of the homogeneous extended SSH model evaluated at the initial momentum $p_0$. This can be seen explicitly in, for example, at $|p_0| = \pi/2$, we have  the numerator becomes $\big( t_1 - (t_2 + t_3) + t_4 \big)^2$ , and $x_{\min}=0$ occurs when $t_1 - (t_2 + t_3) + t_4 = 0$, which is the band closing at $k=\pi$. 
Finally, for a generic critical momentum $p_0 = p_c$, the condition $x_{\min}=0$ requires the numerator of Eq.~\eqref{eq:xmin_extended} to vanish. Since this numerator is a sum of two squares, both terms must vanish simultaneously. This yields $ t_4\,(t_1 - t_4) + t_3\,(t_3 - t_2) = 0$, with $ \cos (2p_c) = \frac{t_3 - t_2}{2 t_4}$, which is precisely the finite-momentum gap-closing condition of the extended SSH chain, with $k_c = 2p_c$.

More generally, Eq.~\eqref{eq:xmin_extended} is just one branch of the turning point obtained by choosing $p=0$ at the turning position. The semiclassical condition $\dot{x} = \partial_p E(x,p) = 0$ can, however, be satisfied at other values of $p$ as well. For example, at $p = \pi/2$ the group velocity of the homogeneous extended SSH band also vanishes, and repeating the same steps gives an analogous expression for $x_{\min}$ in which the denominator is the squared band energy at $k=\pi$, namely $(t_1 - t_2 - t_3 + t_4)^2$, while the numerator remains the same function of $p_0$ as in Eq.~\eqref{eq:xmin_extended}, which one might use in case of $(t_1 + t_2 + t_3 + t_4) = 0$. at $p_0 = 0$, in this expression, the numerator reduces to $(t_1 + t_2 + t_3 + t_4)^2 $, and we obtain $x_{\min} = 0$ when
    $( t_1 + t_2 + t_3 + t_4) = 0$,
which is the band closing at $k=0$.  Thus, all the bulk gap closing conditions at Eq.\ref{eq:constraints} of the extended SSH model can be recovered consistently from the turning point expression by demanding that the numerator of Eq.~\eqref{eq:xmin_extended} vanishes at suitable initial momenta $p_0$. In the present work, we use Eq.~\eqref{eq:xmin_extended} as a convenient representative branch to illustrate how the extended CST--SSH dynamics encodes the three bulk gap closing constraints through the dependence of $x_{\min}$ on the hopping amplitudes and the initial momentum $p_0$.

Thus, we find that the gap-closing conditions for homogeneous extended SSH models, which determine the topological transitions between different winding numbers, also correspond to the conditions under which $x_{\min}$ becomes zero in the extended CST-SSH model. This indicates that horizon-like critical slowdown in the extended CST-SSH model emerges only at those parameter values where topological phase transitions occur in the homogeneous extended SSH model.
 \begin{figure}
     \centering
     \includegraphics[width=.4\textwidth]{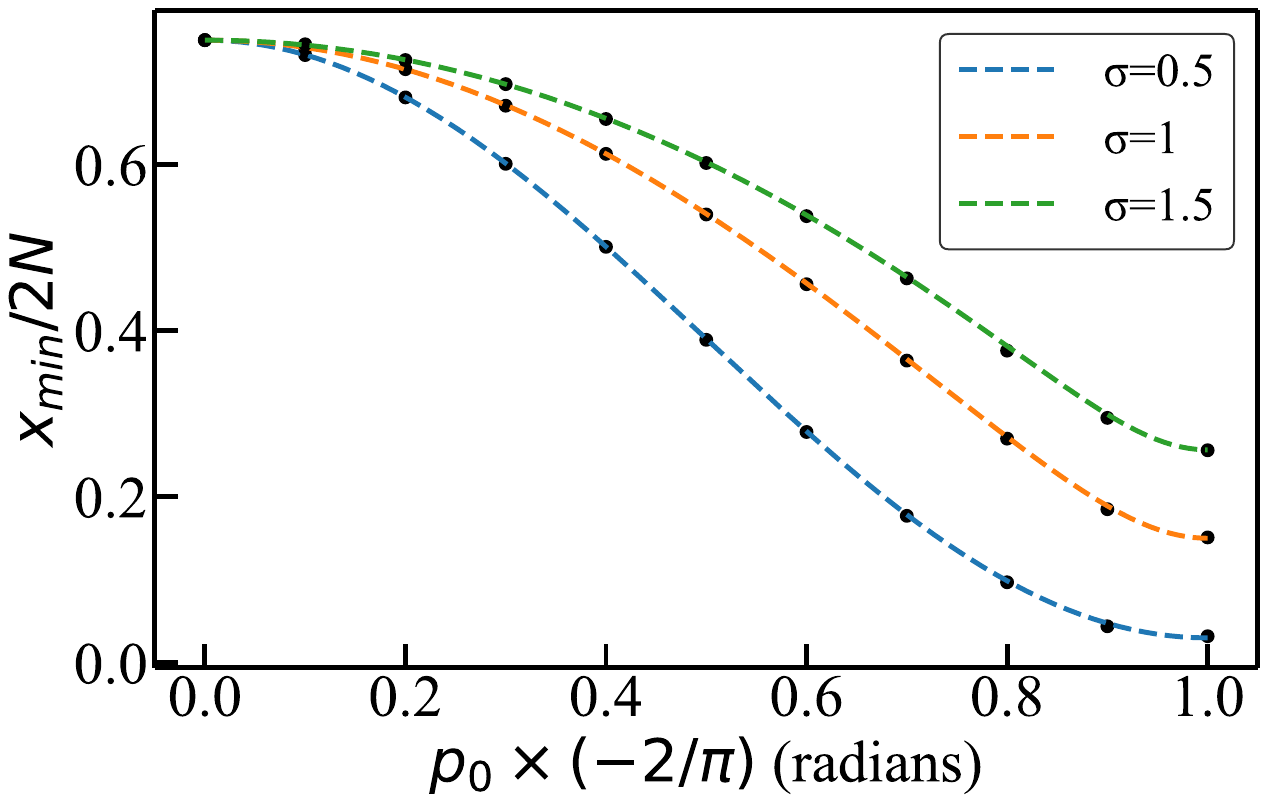}\\[-4cm]
     \caption{ Plotting of turning point by numerical calculation ( black dotted lines) and from semiclassical formula \ref{eq: turning point} (colored dashed lines) at different initial momenta $p_0$ for different values of $\sigma$, at $(t_1,t_2,t_3,t_4)=(1,1.5,0,0)$, for system size $2N=1000$ (500 unit cells), }
     \label{fig:turning point}
 \end{figure}

\captionsetup[subfloat]{position=top, skip=-5pt}

\begin{figure*}[!t]
  \centering
  
  \subfloat[]{%
    \includegraphics[width=0.3\textwidth]{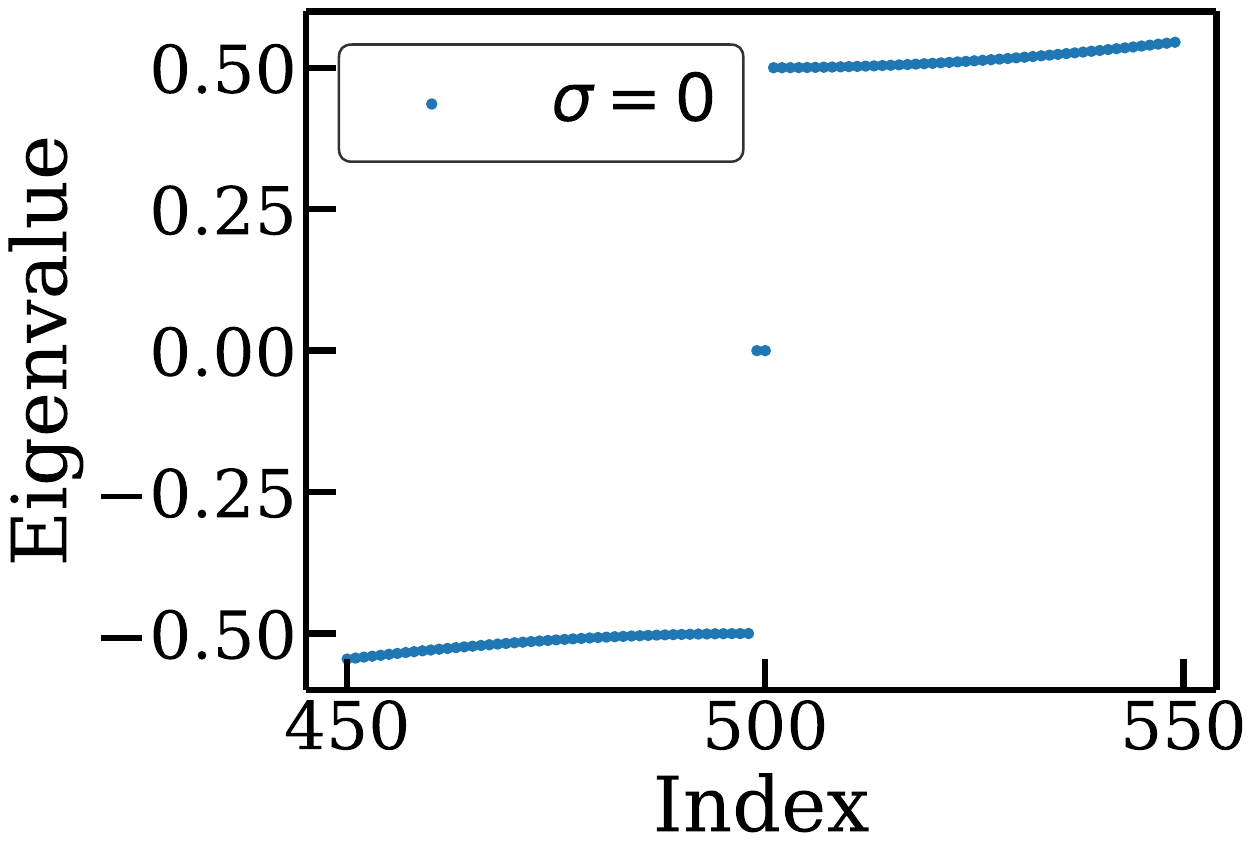}%
    \label{fig:sigma1_eig}}%
  \hfill
  \subfloat[]{%
    \includegraphics[width=0.3\textwidth]{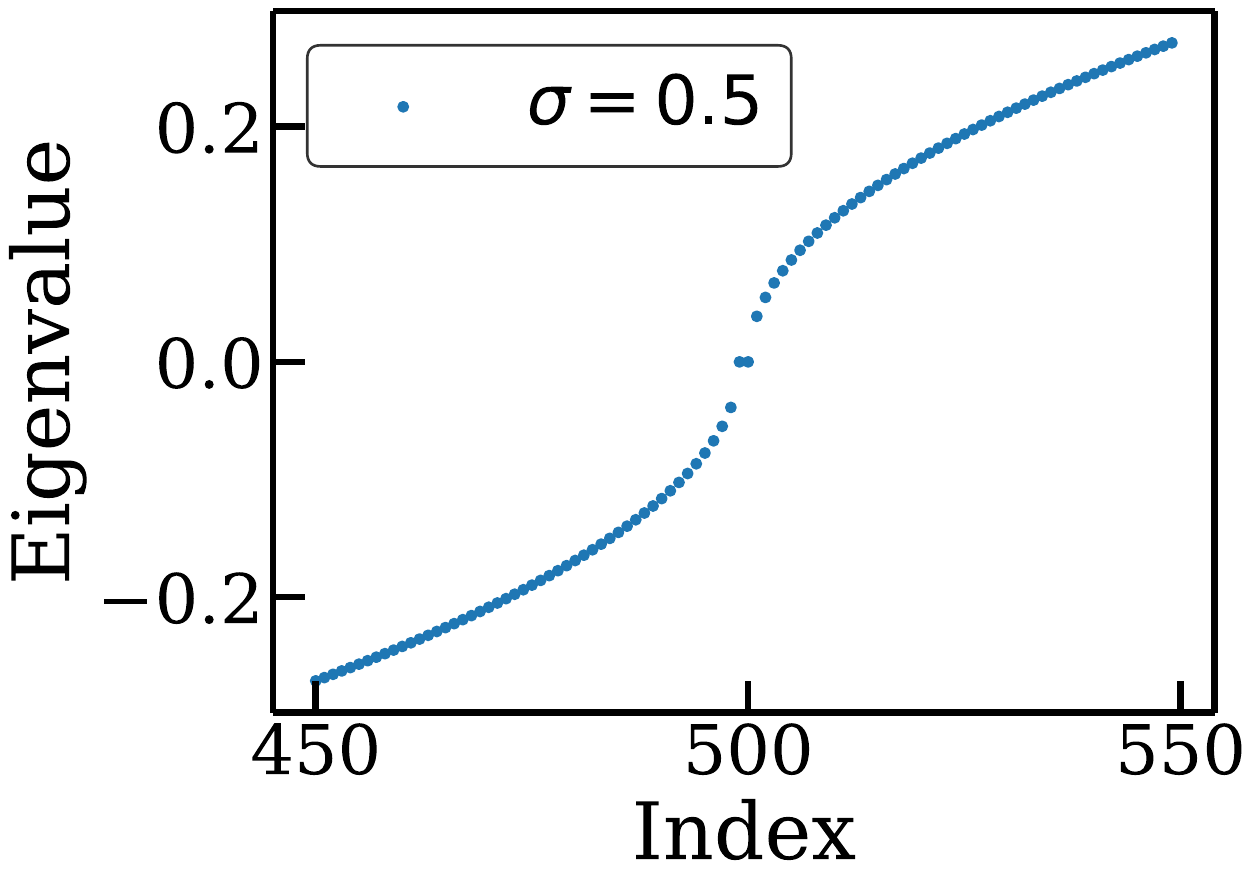}%
    \label{fig:sigma2_eig}}%
  \hfill
  \subfloat[]{%
    \includegraphics[width=0.3\textwidth]{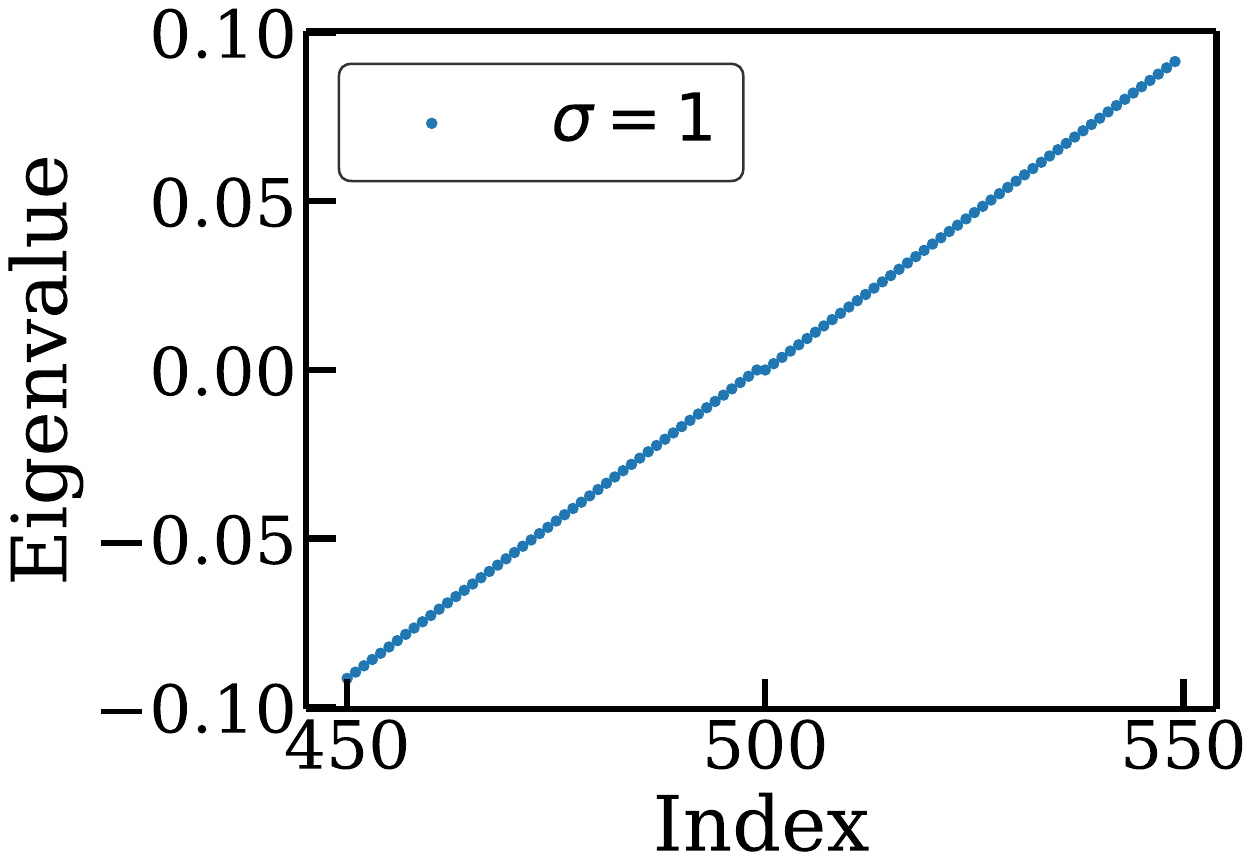}%
    \label{fig:sigma3_eig}}\\[-3cm]

  \subfloat[]{%
    \includegraphics[width=0.3\textwidth]{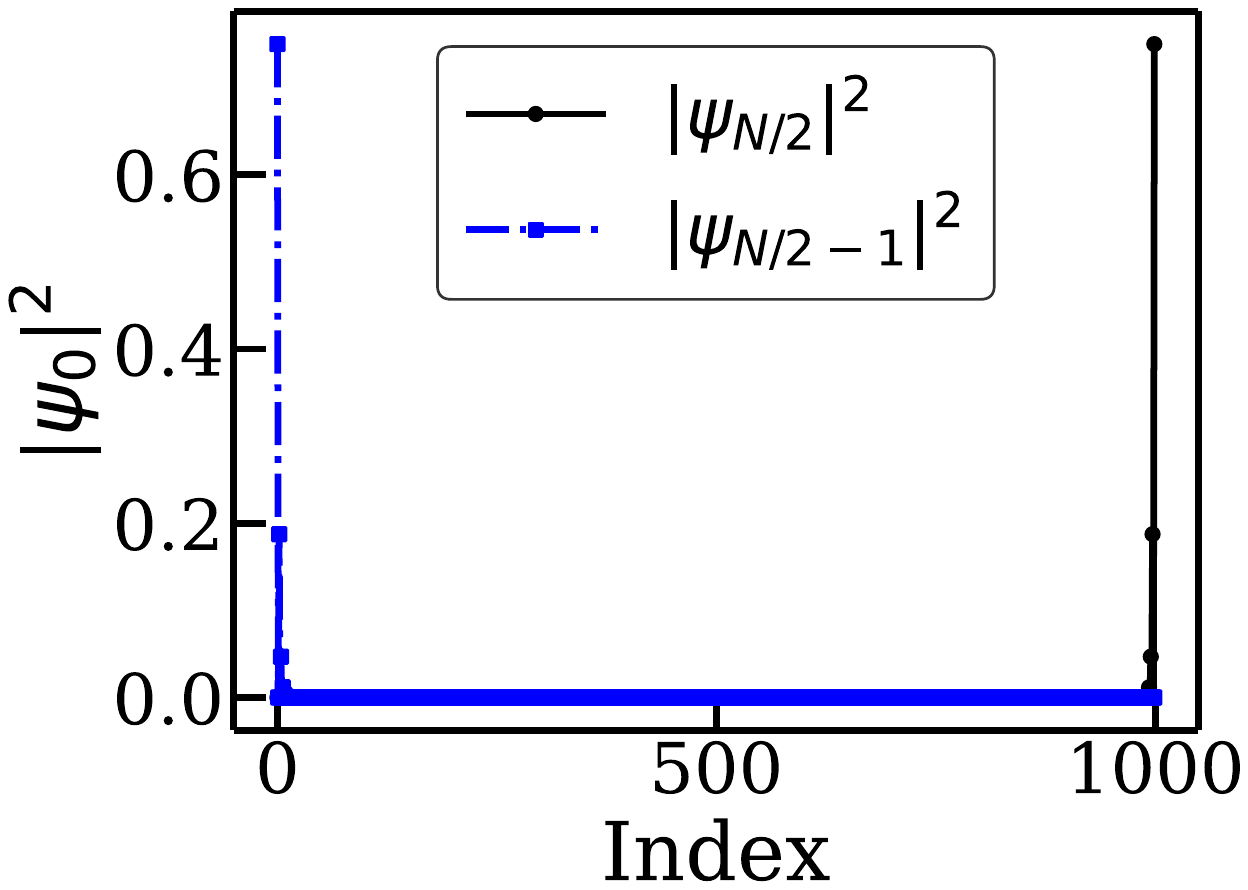}%
    \label{fig:sigma1_state}}%
  \hfill
  \subfloat[]{%
    \includegraphics[width=0.3\textwidth]{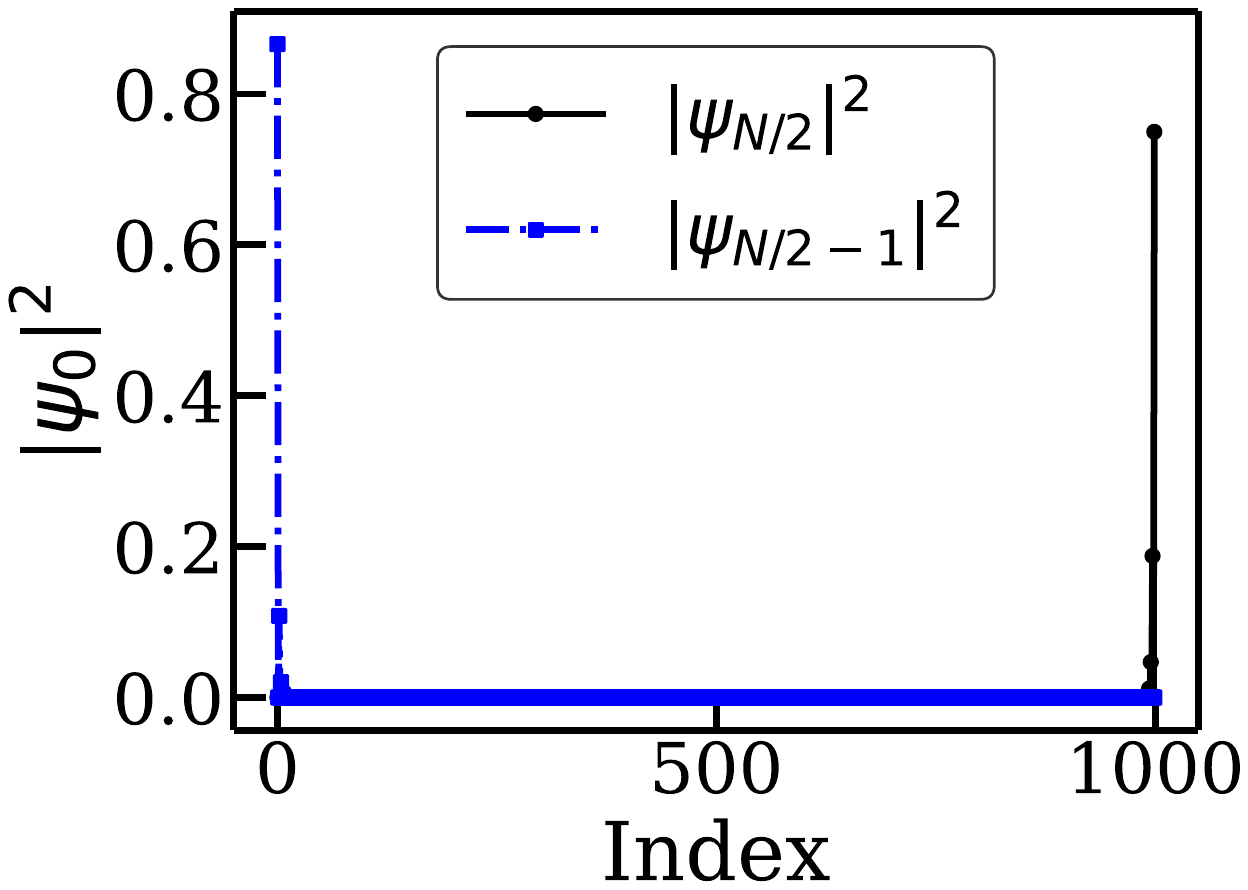}%
    \label{fig:sigma2_state}}%
  \hfill
  \subfloat[]{%
    \includegraphics[width=0.3\textwidth]{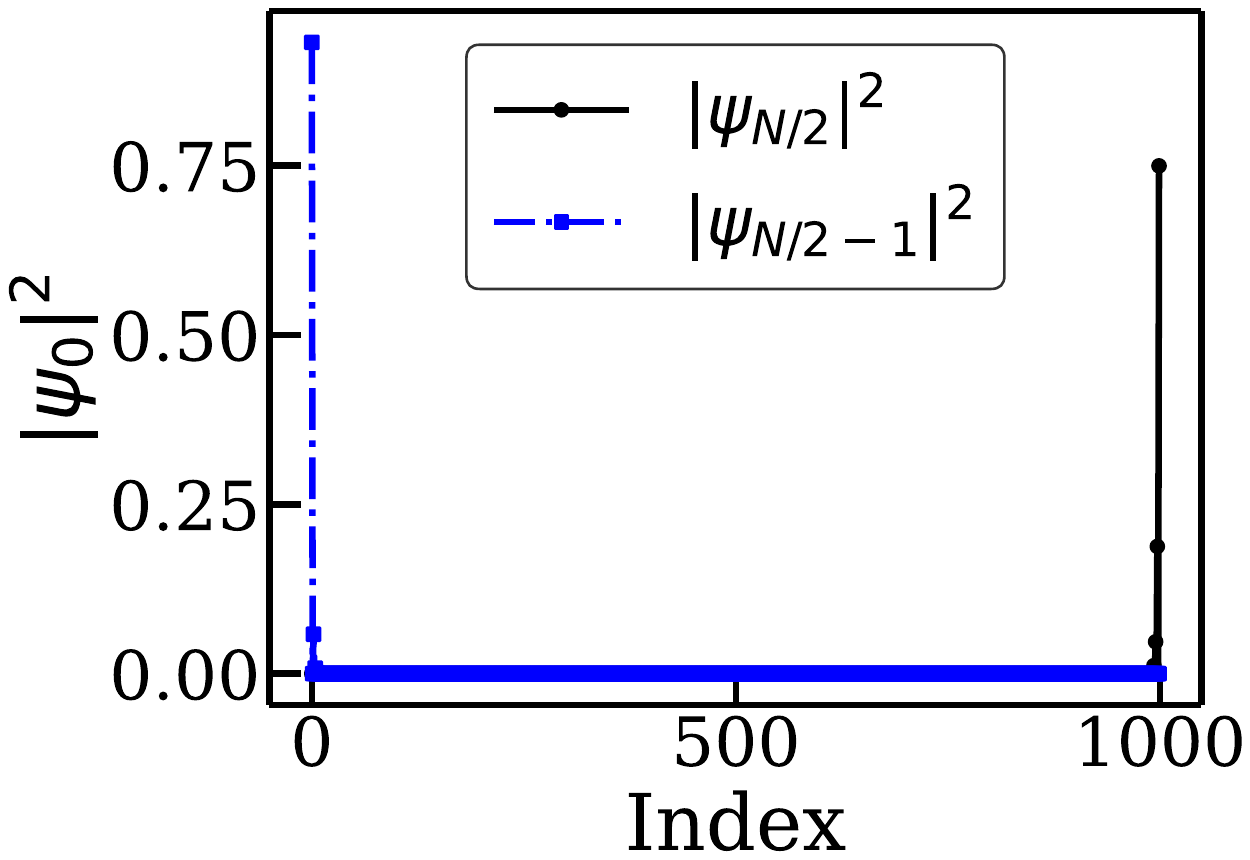}%
    \label{fig:sigma3_state}}\\[-2.5cm]
  
  \caption{Closing of the energy band gap and variation of zero energy eigen states variation with $\sigma$: for a system size $2N=1000$ (500 unit cells) with $(t_1,t_2,t_3,t_4)=(0.5,1,0,0)$ . Top row  (a)--(c): Eigenvalue spectra for $\sigma=0,0.5$, and $1$ respectively. Bottom row (d)--(f): Corresponding zero-energy eigenstates profiles.}
  \label{fig:zero energy states vs sigma}
\end{figure*}

In a similar spirit to the previous calculations, we can analyze the semiclassical dynamics for the CST version of the standard SSH Hamiltonian, the limiting case of Eq.~\ref{eq:ext_CST_SSH} with $t_3=t_4=0$. After obtaining the recursive energy eigenvalue equations and the continuum Hamiltonian, we can obtain the semiclassical equation of motion as:

\begin{equation}
\dot{x}
\;=\;
\frac{\partial E_{\pm}}{\partial p}
\;=\;
\mp\,
\frac{\,2t_{1}(x)\,t_{2}(x)\sin\bigl(2p\bigr)\,}{
\sqrt{\,t_{1}^2(x) + t_{2}^2(x) + 2\,t_{1}(x)\,t_{2}(x)\,\cos(2p)\,}
}
\end{equation}

\begin{equation}
\begin{split}
&\dot{p}
\;=\;
-\frac{\partial E_{\pm}}{\partial x}
\\
&=\mp
\,
\frac{
t_{1}(x)t_{1}'(x)
+
t_{2}(x)t_{2}'(x)
+
\cos(2p)\bigl(t_{1}'(x)\,t_{2}(x) + t_{1}(x)t_{2}'(x)\bigr)
}{
\sqrt{t_{1}^2(x) + t_{2}^2(x) + 2t_{1}(x)t_{2}(x)\cos(2p)\,}
}
\end{split}
\end{equation}
In Figs.~ \ref{fig:sc vs num} and \ref{fig:sc vs num ratio}, the semi-classical trajectories are compared with the peak position of exact dynamics results. Moreover, the semiclassical result can also mimic the event horizon like critical slowdown for $t_1=t_2$ and $p_0=-\pi/2$. With $t_1=t_2$, the Hamiltonian matrix essentially becomes identical with the CST-TB Hamiltonian  \cite{morice2021synthetic}, in other words, it can also be referred to as the topological transition point of the usual SSH model. Semiclassical equations can similarly allow us to predict the turning point $x_{\min}$ for $p_0\neq -\pi/2$  
, 

 \begin{equation}
     \begin{split}
        & x_{\min}=
         x_{0}\Big[\frac{t_{1}^2 \;+\; t_{2}^2  \;+\; 2\,t_{1}\,t_{2}\,\cos\!\bigl(\,2p_0\bigr)}{(t_1+t_2)^2}\Big]^{\frac{1}{2\sigma}}
     \end{split}
     \label{eq: turning point}
 \end{equation}
 Fig.~\ref{fig:turning point} shows how the turning point varies with different $p_0$ values at different $\sigma$. The numerical and analytical turning points are extremely well lined up, solidifying the validity of semiclassical calculations. Moreover, the equation suggests that for $x_{\min}$ to be zero, the two hopping $t_1$ and $t_2$ must be equal, leading to a more simplified expression $x_{\min}=x_0(\cos p_0)^{\frac{1}{\sigma}}$, when $p_0=\pm\pi/2$, only then $x_{\min}$ becomes $0$. Hence, in the case of $t_1 \neq t_2$, the wave packet has no initial momentum $p_0$ value for which the turning point becomes $0$. It also validates our previous numerical finding, i.e., only when $t_1=t_2$, the model can show the horizon like critical slowdown, and that too for $p_0=-\pi/2$, which also has been observed for the tight binding lattice model~\cite{morice2021synthetic, morice_2022}.

\section{Analytical expression for the exact zero energy eigenfunctions\label{gap}}
From the wave-packet dynamics, it is apparent that the extended CST-SSH model can also mimic
horizon like critical slowdown for certain choices of parameters. This naturally raises the question of whether
this model also exhibits a topological phase transition. 
Note that in the $\sigma=0$ limit, this model reduces to the usual extended SSH model, which
undergoes a topological phase transition between a topologically trivial phase and a
topologically non-trivial phase. In the topologically non-trivial phase, under open-boundary
conditions, zero-energy states appear that are localized at the two edges of the system.
An immediate question then arises: what happens to these edge states when $\sigma>0$?
In particular, do such edge states survive in the presence of the CST deformation?
In the following, we analyze this question. For analytical simplicity, we restrict our
discussion to the CST-SSH model; however, the analysis can be straightforwardly extended
to the extended CST-SSH model.
First, we perform a numerical check in Fig.~\ref{fig:zero energy states vs sigma}. Figures in the lower panel suggest that zero energy states still survive for finite $\sigma$, though the upper panel figures showing gap is becoming smaller with increasing $\sigma$. Moreover, those states remain localized at the edges. However, unlike the usual SSH model, the states are asymmetric; the left edge state seems to become more localized compared to the right edge state with increasing $\sigma$.   

\begin{figure}
     \centering
     \includegraphics[width=.36\textwidth]{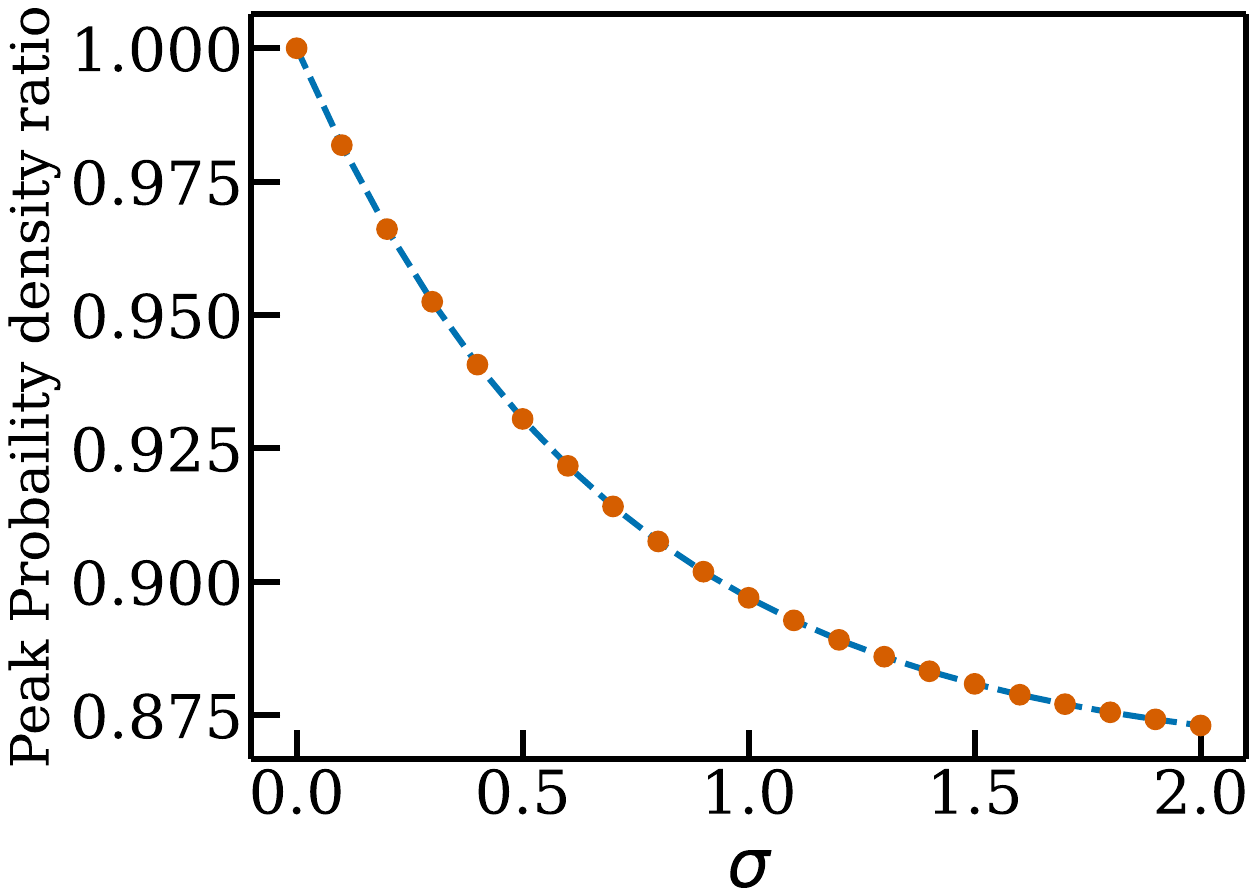}\\[-3.5cm]
     \caption{Ratio of  right-to-left peak of probability densities vs $\sigma$ for a system size $2N=1000$ (500 unit cells) with $(t_1,t_2,t_3,t_4)=(0.5,1,0,0)$. Here the dotted lines represent the numerical calculations and the dashed lines are from the analytical expressions of the zero energy eigenstates \ref{eq:left edge zes},\ref{eq:right edge zes}}
     \label{fig:peak ratio}
 \end{figure}

It is reasonably straightforward to obtain those zero eigenstates by solving the time-independent Schrödinger equation for the CST-SSH Hamiltonian, and one can easily obtain two recursive relations for two sub-lattices A and B. 
\\ For  sublattice A:
\begin{equation}
    t_2(n-1)\psi_{B,n-1}+t_1(n)\psi_{B,n}=0
    \label{eq: Sublattice A}
\end{equation}
\\For sublattice B:
\begin{equation}
    t_1(n)\psi_{A,n}+t_2(n)\psi_{A,n+1}=0
    \label{eq: Sublattice B}
\end{equation}
Now, from \eqref{eq: Sublattice A} ,
\begin{equation}
   \begin{split}
\psi_{B,n} &= -\frac{t_1(n+1)}{t_2(n)}\psi_{B,n+1}\\
           &= \Big(-\frac{t_1(n+1)}{t_2(n)}\Big)\Big(-\frac{t_1(n+2)}{t_2(n+1)}\Big)\cdots\Big(-\frac{t_1(N)}{t_2(N-1)}\Big)\psi_{B,N}\\
           &=\big(-\frac{t_1}{t_2}\big)^{N-n}\prod_{m=n+1}^{N}\Big( \frac{2m-1}{2(m-1)}\Big)^\sigma\psi_{B,N}
           \label{eq:right edge zes}
\end{split}
\end{equation}
\\Likewise, for sublattice B, we can obtain from \eqref{eq: Sublattice B},
\begin{equation}
    \begin{split}
        \psi_{A,n} &=-\frac{t_1(n-1)}{t_2(n-1)}\psi_{A,n-1} = \Big( -\frac{t_1}{t_2} \Big)^{n-1}\prod_{m=1}^{n-1}\Big(\frac{2m-1}{2m}\Big)^{\sigma}\psi_{A,1}
    \end{split}
    \label{eq:left edge zes}
\end{equation}
These two zero-energy eigenstates $\psi_{A,n}$ and $\psi_{B,n}$ will appear on the left and right edges of the chain, respectively. $\psi_{B,N}$ and $ \psi_{A,1}$ can easily be obtained by normalizing the eigen states. For the standard SSH model, with $\sigma=0$, both states are identical, only one is peaked at the left edge, and the other at the right. But with the introduction of $\sigma$, asymmetry arises. The state on the right edge remains almost the same for non-zero $\sigma$, and the peak height on the left edge keeps increasing with $\sigma$.  
The above results can also be understood intuitively. For $\sigma \ge 0$, the hopping amplitudes increase monotonically from the left to the right end of the chain via the factor $\sim (n/N)^\sigma$. Consequently, hopping near the left edge is strongly suppressed, leading to enhanced localization of the left-edge state with reduced bulk penetration. In contrast, near the right edge $(n/N)^\sigma \approx 1$, and the edge state behaves similarly to that of the standard SSH model. Fig.~\ref{fig:peak ratio} shows how the right-to-left peak ratio of probability densities varies with $\sigma$, showing exact matching with the ratio calculation from the analytical expression.

\begin{figure}
  \centering
  
  \subfloat[]{%
    \includegraphics[width=0.41\textwidth]{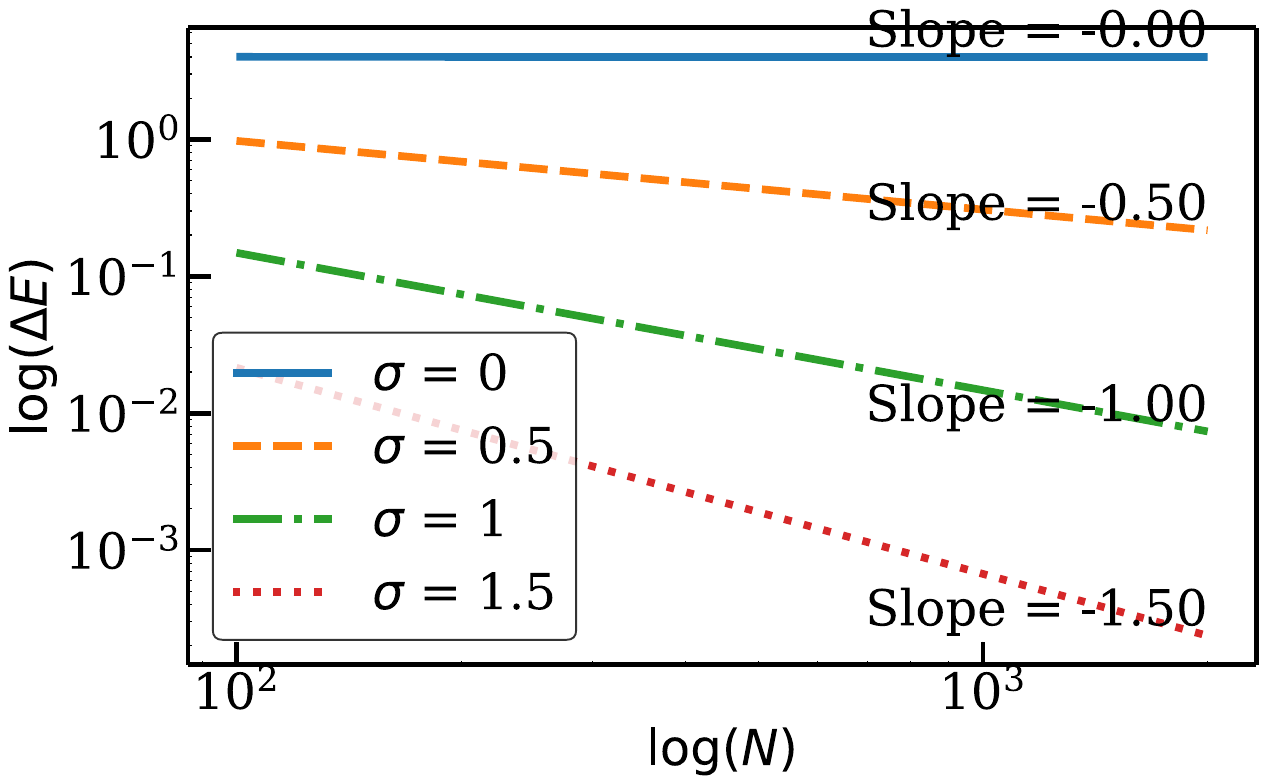}%
    \label{fig:gap_closing_extended_w2}}%
    \hfill
   \\[-4cm]
  \subfloat[]{%
    \includegraphics[width=0.41\textwidth]{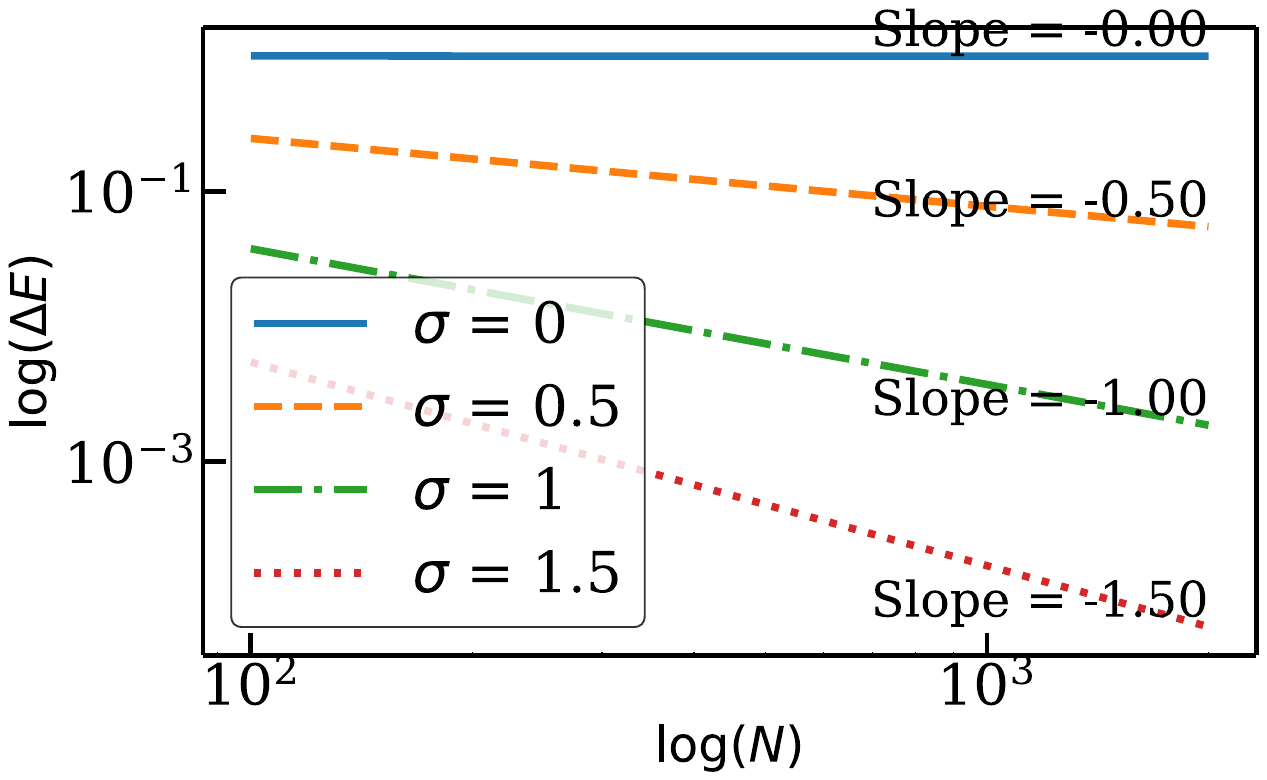}%
    \label{fig:energy_gap_vs_N_log_log}}\\[-4cm]
   
  \caption{ Energy gap ($\Delta E$) at various system size $N$ for different $\sigma$, for the Extended CST-SSH with (a) $(t_1,t_2,t_3,t_4)=(2,1,-1,-4)$ , and, for (b) $(t_1,t_2,t_3,t_4)=(0.5,1,0,0)$}
  \label{fig:gap_closing}
\end{figure}

Another thing to note is the closing of the energy gap ($\Delta E$) with $\sigma$. We observe that the gap gradually closes with increasing $\sigma$ and becomes almost non-existent as $\sigma$ reaches $1$. Such gaplessness in topological insulator due to inhomogeneity or disorder has also been observed in the literature ~\cite{mobility_gap_2012,mondal2025topologicalphasetransitioninfinite}. From Fig.~\ref{fig:gap_closing}, we observed that $\Delta E$ decreases with $\sigma$ as $\Delta E \propto N^{-\sigma}$ for a  fixed $t_1, t_2, t_3, t_4$. 
Note that the finite-size system on the lattice will always have a gap, and a typical gap between two adjacent energy eigenvalues decreases with system size as $N^{-1}$ (we have found the same for our model as well). 
Thus, the ratio between $\Delta E$ and the typical gap in the spectrum scales as  $N^{1-\sigma}$, indicating the presence of isolated zero-energy states even for non-zero $\sigma<1$. The fact that $N\Delta E$ increases with $N$, suggests the system remains gapped even in the thermodynamic limit for $\sigma<1$~\cite{PhysRevB.84.115135}, and gapless for $\sigma\geq1$ 
(see Fig.~\ref{fig:gap_closing} and also Appendix \ref{Appendix_ext_CST_SSH}). 
This observation indicates that for any combination of $t_1, t_2, t_3, t_4$, even if the parameters
do not correspond to a topological phase transition point, the system becomes gapless in the
$\sigma \geq 1$ region. However, in this regime we do not observe horizon-like critical slowdown. Hence, the
presence of a gapless spectrum alone does not guarantee the occurrence of a critical slowdown.
Furthermore, unlike the standard SSH model, the extended SSH model cannot, at the
topological transition points, be mapped onto a single-band tight-binding CST Hamiltonian. Nevertheless, we also observe well-localized zero-energy edge states for the $\sigma$ higher than $1$ as well.  

\section{Topological signatures\label{signature}}
In the previous section, the survival of the zero-energy edge states for $\sigma>0$ shows the
signature that 
 the CST version of the SSH model can retain the imprint of the same topological
phases and transition of the usual SSH models. However, the CST-SSH models remain gapless for $\sigma\geq1$. In order to 
understand that further, one needs to understand the symmetries of the model, as well as some proper topological markers, which are required. First, we investigate whether even after the introduction of warping of spacetime $(\sigma)$, the Hamiltonian will still remain in the BDI category with winding number as its topological invariant quantity, as it is observed for the usual SSH model. Since the position-dependent hopping terms are real, the time reversal operator acting on the Hamiltonian $\hat{T}H\hat{T}^{-1}= H$ leaves it invariant. Moreover, the Hamiltonian being a spinless fermionic system,  $\hat{T}^2=\mathbb{I}$. For chiral symmetry, first, we define $\hat\Gamma_A= \sum_{n=1}^Nc_{n,A}^\dagger c_{n,A}$ and $\hat\Gamma_B= \sum_{n=1}^Nc_{n,B}^\dagger c_{n,B}$ as projectors on sublattice A and B~\cite{Asb_th_2016}. Then, the chiral operator is defined by $\hat\Gamma=\hat\Gamma_A-\hat\Gamma_B$, which anti-commutes with the Hamiltonian $\hat\Gamma H\hat\Gamma=-H$, since the bipartite nature is maintained in the modified position-dependent SSH models. This anti-commutation relation holds regardless of the position-dependent nature of the hopping amplitudes~\cite{Asb_th_2016}. Time reversal and Chiral symmetry being present with $T^2=\Gamma^2=\mathbb{I}$, we can combine them as, $\hat{C}=\hat{T} \hat{\Gamma}$, which can acts as a particle hole symmetry, i.e., an anti-unitary operator that anti-commutes with the Hamiltonian, and also $\hat{C}^2=\mathbb{I}$. Thus, the CST-SSH Hamiltonians remain in the BDI category in the  Altland-Zirnbauer tenfold classification of topological insulators and superconductors \cite{Altland_1997,Ryu_2010} with an integer $\mathbb{Z}$ topological invariant like its standard counterpart in the usual SSH model. Therefore, the symmetries remain invariant under the position-dependent hopping amplitude, and so does the topological invariant, i.e., winding number. However, in the CST-SSH models, the absence of translational symmetry prevents us from utilizing the k-space to calculate the winding number. Therefore, we adopt two real space markers to calculate the topological invariants: (i) Local topological marker (LTM),
(ii) Mean chiral displacement (MCD). 

\subsection{Local Topological Marker\label{marker}} 

\begin{figure}
  \centering
  
  \subfloat[]{%
    \includegraphics[width=0.4\textwidth]{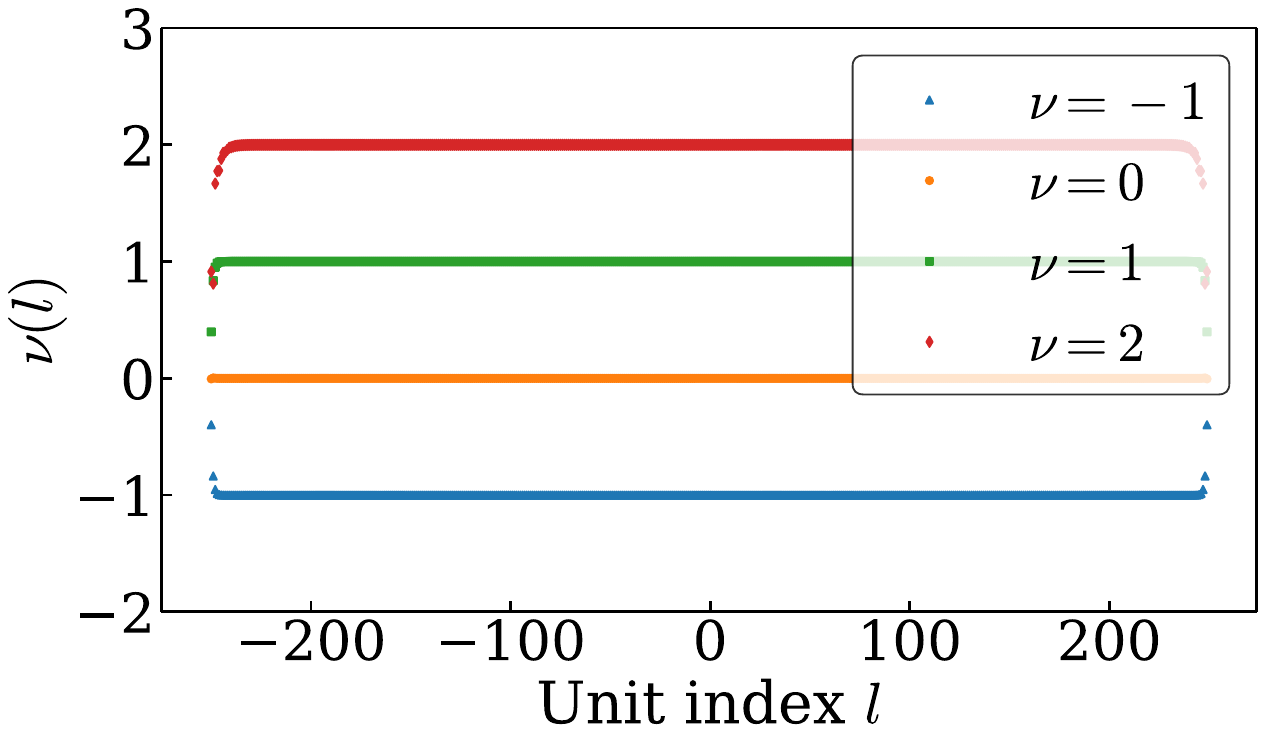}%
    \label{fig:ext_LTM_sigma0}}%
    
  \hfill
  \\[-4cm]
  \subfloat[]{%
    \includegraphics[width=0.4\textwidth]{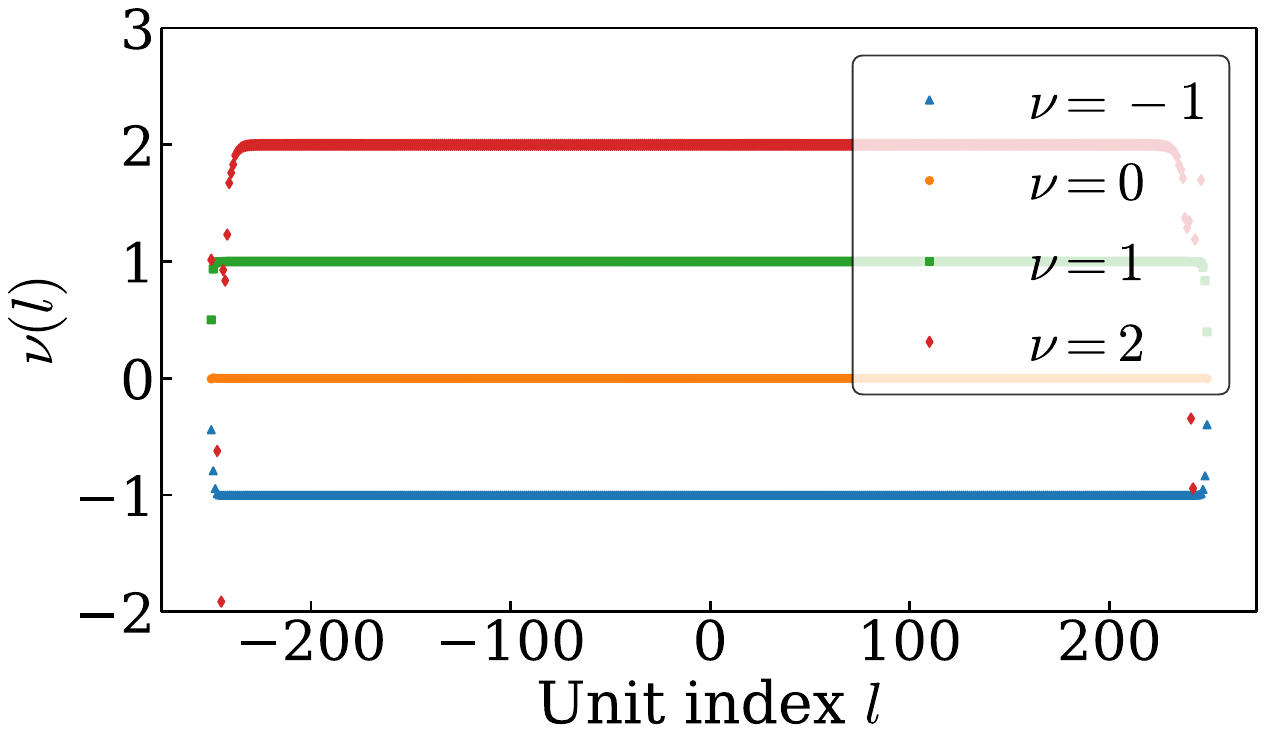}%
    \label{fig:ex_LTM simga1}}\\[-4cm]
   
  \caption{Local Topological Marker (LTM) for extended CST-SSH model $(a) \sigma=0$ and $(b) \sigma=1$ , for the CST-SSH model having system size $2N=1000$ (500 unit cells), giving the same winding numbers for both $\sigma$ values. $(t_1,t_2,t_3,t_4)=(0.5, 0, 1, 0), (1,0.5,0,0),(0.5,1,0,0), (0,1,1,-3)$ are chosen to obtain winding numbers $-1,0,1,2$, respectively.}
  \label{fig:ext_LTM}
\end{figure}

   \begin{figure}
  \centering
  \subfloat[]{%
    \includegraphics[width=0.4\textwidth]{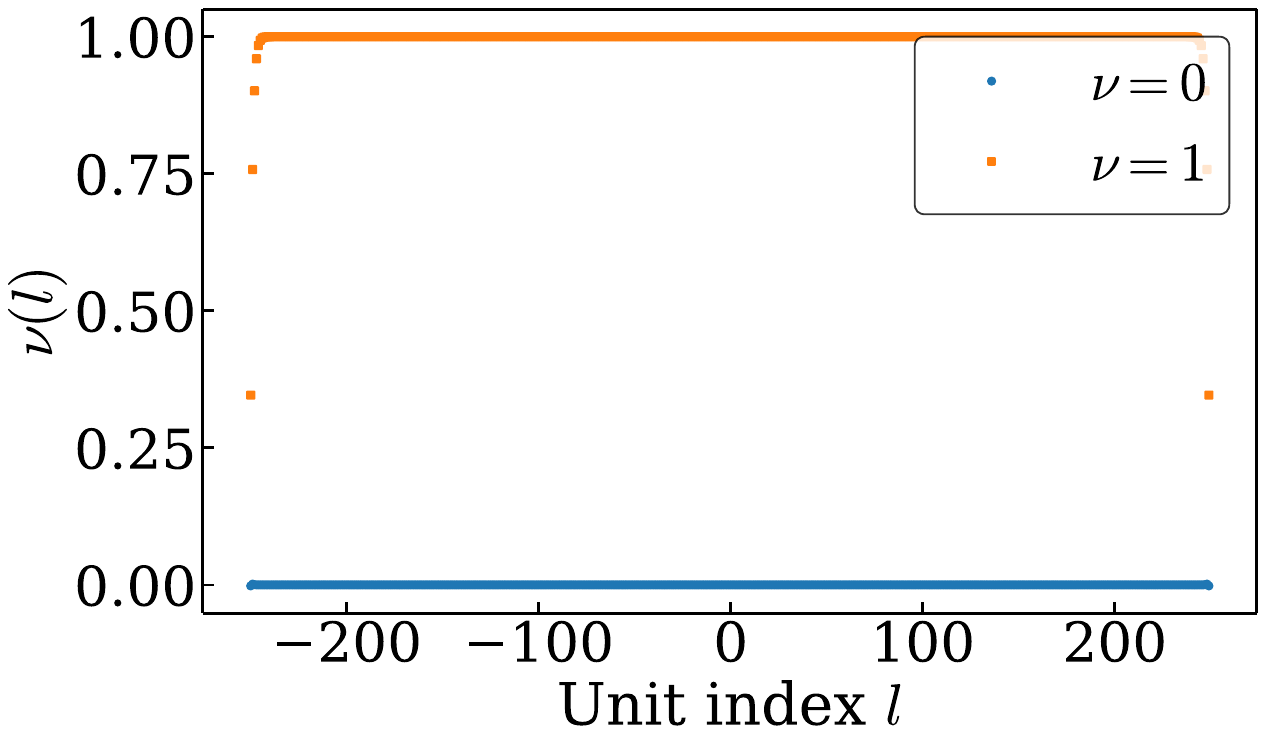}%
    \label{fig:LTM_sigma0}}%
  \hfill
  \\[-4cm]
  \subfloat[]{%
    \includegraphics[width=0.4\textwidth]{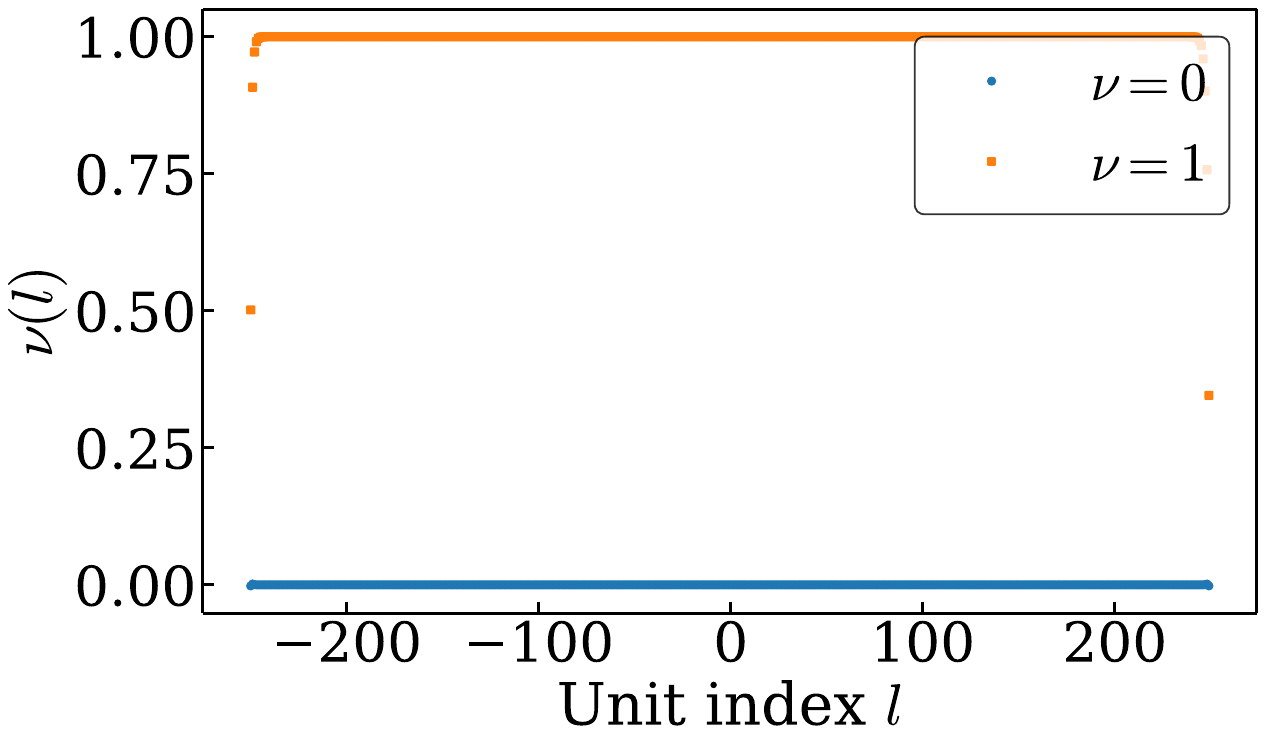}%
    \label{fig:LTM simga1}}\\[-3.5cm]
  \caption{Local Topological Marker (LTM) for $(a) \sigma=0$ and, $(b) \sigma=1$ for trivial $(t_1,t_2,t_3,t_4)=(1,0.4,0,0)$ and topological phase  $(t_1,t_2,t_3,t_4)=(0.6,1,0,0)$, for the CST-SSH model having system size $2N=1000$ (500 unit cells) }
  \label{fig:LTM}
\end{figure}

\begin{figure}
  \centering
  \subfloat[]{%
    \includegraphics[width=0.4\textwidth]{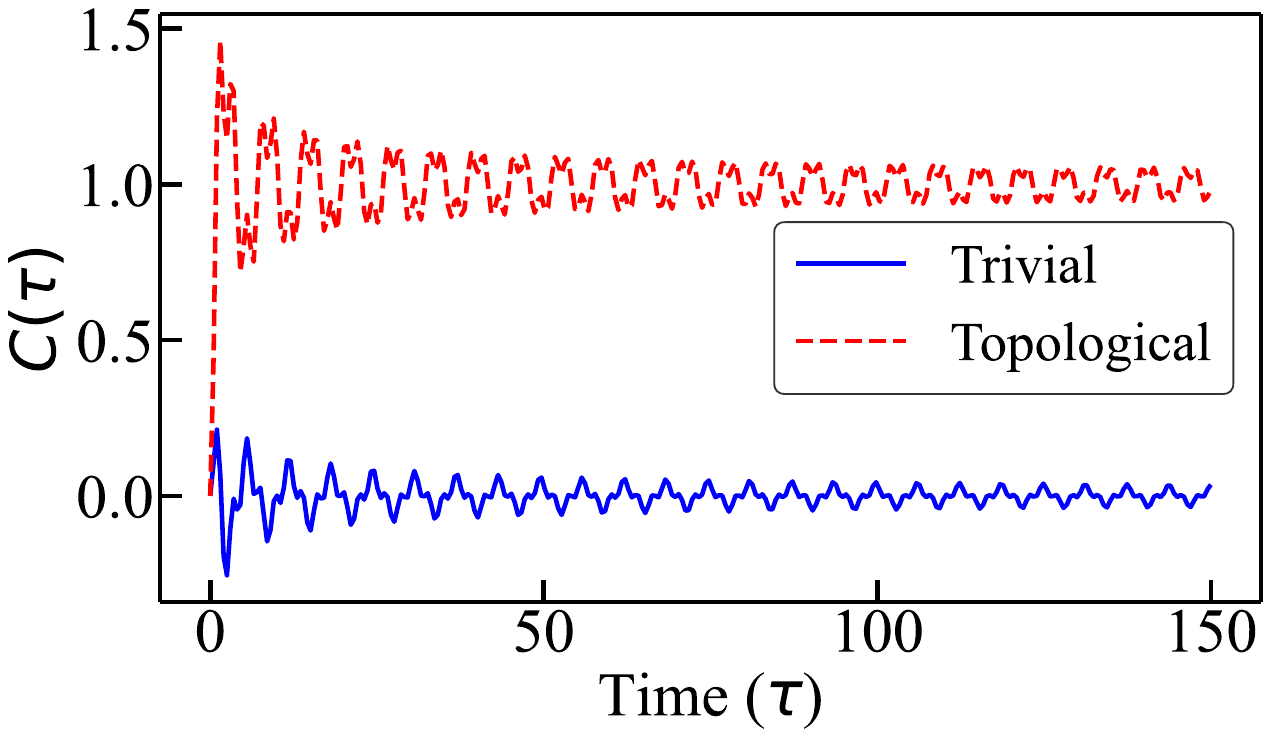}%
    \label{fig:MCD_sig_0}}%
  \hfill
  \\[-4cm]
  \subfloat[]{%
    \includegraphics[width=0.4\textwidth]{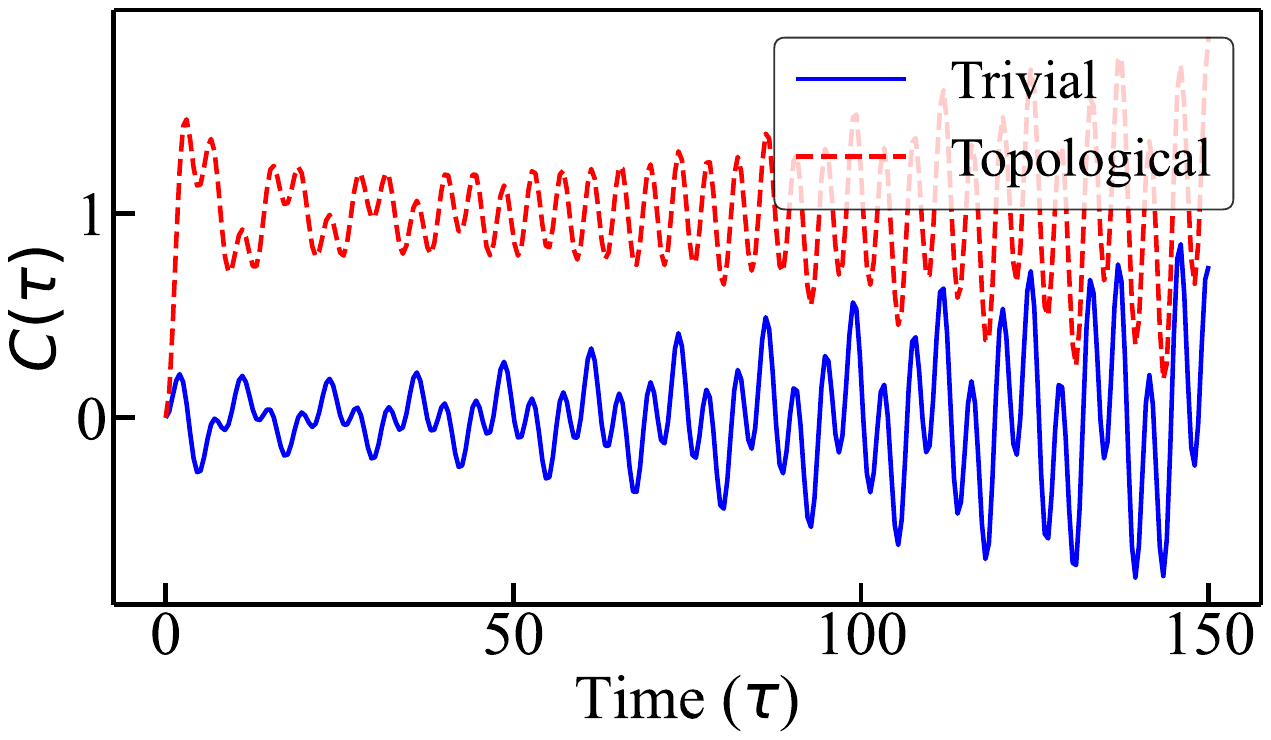}%
    \label{fig:MCD_sig_1}}\\[-4cm]
  \caption{Time dependent MCD for both topological (red) with  $(t_1,t_2,t_3,t_4)=(0.5,1,0,0)$ and trivial (blue) with $(t_1,t_2,t_3,t_4)=(1,0.5,0,0)$ for (a) $\sigma=0$ and,(b) $\sigma=1$ for system size $2N=1000$ (500 unit cells)  }
  \label{fig:MCD}
\end{figure}

For a dimer chain with broken translational symmetry, the Local Topological Marker (LTM) can give the value of the topological invariant quantity in its bulk. We here obtain the winding number in real space using LTM~\cite{PhysRevResearch.3.033012, ai_wire}. It is based on the rearranged eigenfunctions corresponding to the ascending eigenvalues. The LTM can be defined as:

\begin{equation}
\nu(l)=\frac{1}{2}\sum\limits_{a=A,B}\{(\mathbb Q_{BA}[\mathbb X,\mathbb Q_{AB}])_{la,la}+(\mathbb Q_{AB}[\mathbb Q_{BA},\mathbb X])_{la,la}\},
    \label{LTM}
\end{equation}
     Here, $\mathbb X$ is the position operator.  $\mathbb Q$ can be defined from the modal matrix $\mathbb{U}$, which is comprised of all the normalized eigen vectors with ascending order, explicitly, $\mathbb U= [\mathbf U_1,\mathbf U_2,...\mathbf U_n,\mathbf U_{n+1},...\mathbf U_N]$. Here, $\mathbb U_-=[\mathbf U_1,\mathbf U_2,...\mathbf U_n],$ and  $\mathbb U_+-=[\mathbf U_{n+1},\mathbf U_{n+2},...\mathbf U_N] $ are corresponds to below and above the band gap energy spectrum. 
 Then one can define the projectors as $\mathbb P_-=\mathbb U_-\mathbb U_-^T$ and $\mathbb P_+=\mathbb U_+\mathbb U_+^T$, and $\mathbb Q$ as, $\mathbb Q=\mathbb P_+-\mathbb P_-$, which further can be decomposed as, $\mathbb Q=\mathbb Q_{AB}+\mathbb Q_{BA}= \Gamma_A\mathbb Q\Gamma_B+\Gamma_B\mathbb Q\Gamma_A$, where $\Gamma=\Gamma_A-\Gamma_B$ is the chiral operator. More about the formula and the operators can be found in the appendix of \cite{PhysRevResearch.3.033012, ai_wire}. From the Figs. \ref{fig:ext_LTM}, \ref{fig:LTM}, we observe that even for $\sigma=1$, the LTM shows the same winding number for the both CST-SSH Hamiltonians as its usual SSH counterparts, for the exact values of the hopping. 
 There is no apparent difference between the standard and the position-dependent hopping SSH models for this static marker, also indicating same topological phase boundaries for the CST models as well. 
\subsection{Mean Chiral Displacement}
The previous topological measure was a static measure; in this section, we use a dynamical topological measure, which we call  Mean Chiral Displacement (MCD) \cite{Cardano_2017, PhysRevResearch.3.033012}, can be used to detect the winding number which is defined as, 
\begin{equation}
C(\tau)=2\langle\psi(\tau)|\Gamma\mathbb X|\psi(\tau)\rangle, 
\end{equation}
where  $|\psi(\tau)\rangle=e^{-iH\tau}\ket{\psi_i}$, i.e., $|\psi(\tau)\rangle$ is the time evolved state of an initially localized state at $n=N/2$ unit cell in the A sublattice, i.e., $\ket{\psi_i}$. $\Gamma,~\mathbb X$ are chiral and displacement operators, respectively. In the figure \ref{fig:MCD}, we show the time-dependent MCD for both the trivial and topological phases, where it oscillates and converges to $0$ in the case of a trivial phase and, on the other hand, saturates to $1$ in the case of a topological phase for the usual SSH model. On the other hand, for  $\sigma\neq0$ also, the MCD oscillates around its respective winding number, but the fluctuations are much higher compared to $\sigma=0$, though the average MCD remains the same. 
Overall, the MCD results complement our previous results for the static measure. This indeed proves that,  like the usual SSH models, even the CST-SSH models display topologically distinct phases corresponding to distinct winding numbers for the same conditions on the hopping as the regular SSH models, though the model remains gapless even away from the topological transition points for $\sigma\geq 1$.   

\begin{figure*}[!t]
  \centering
 
  \subfloat[]{%
    \includegraphics[width=0.4\textwidth]{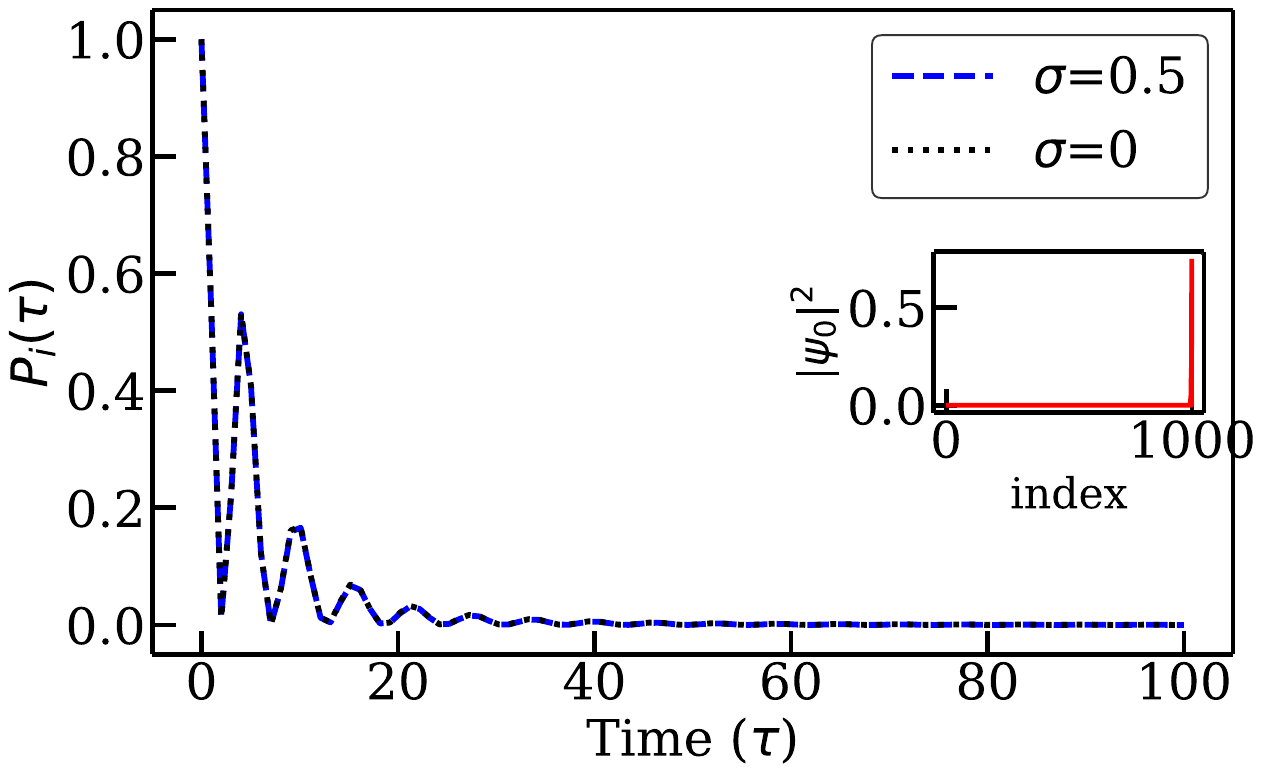}%
    \label{fig:quench_0_to_1_right_sig_0.5}}%
  \hfill
  \subfloat[]{%
    \includegraphics[width=0.4\textwidth]{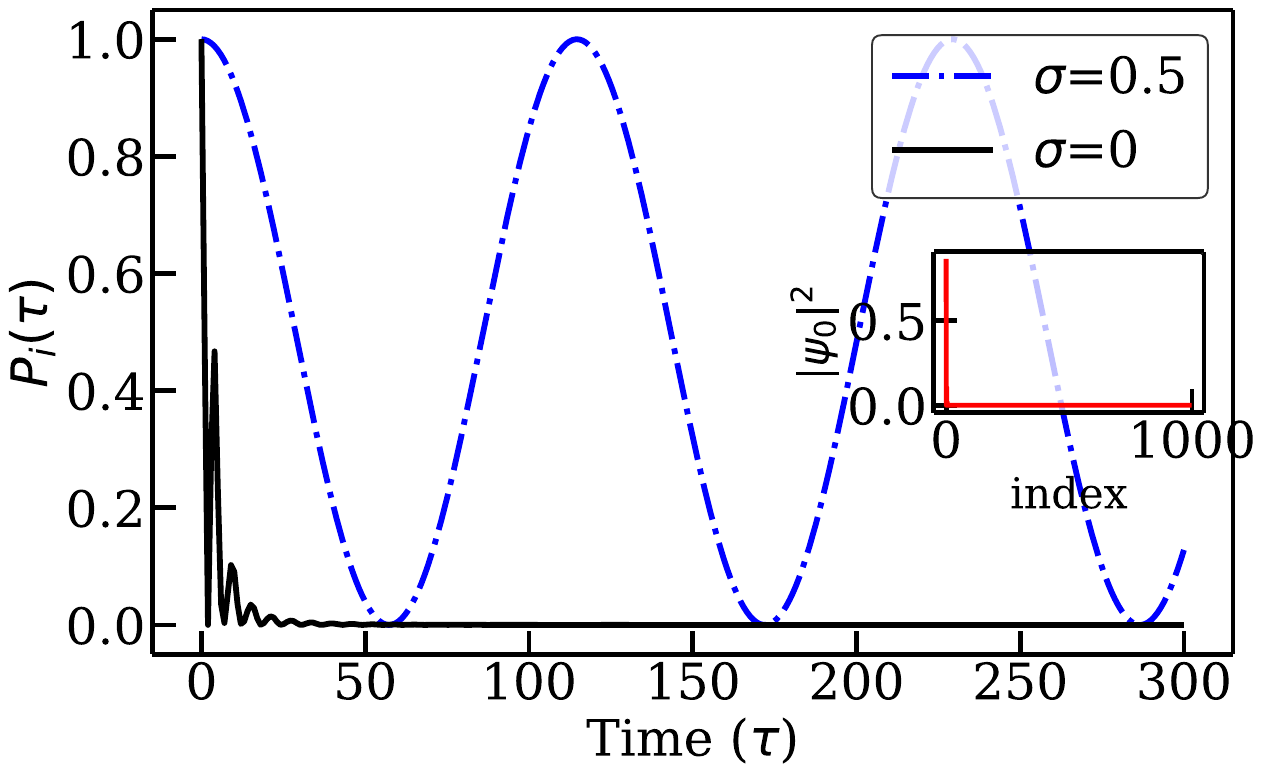}%
    \label{fig:quench_0_to_1_left_sig_0.5}}\\[-5cm]
   
  \caption{Measurement of survival probability $P_i(t)$ from winding number 1 (with $(t_1,t_2,t_3,t_4)=(0.5,1,0,0)$) to winding number 0 (with $(t_1,t_2,t_3,t_4)=(1,0.5,0,0)$) for (a) right edge zero energy eigen state and, (b) left edge zero energy eigen state, for system size $2N=1000$ (500 unit cells)}
  \label{fig:all_sigma}
\end{figure*}
 
 \begin{figure*}[!t]
  \centering
  \subfloat[]{%
    \includegraphics[width=0.42\textwidth]{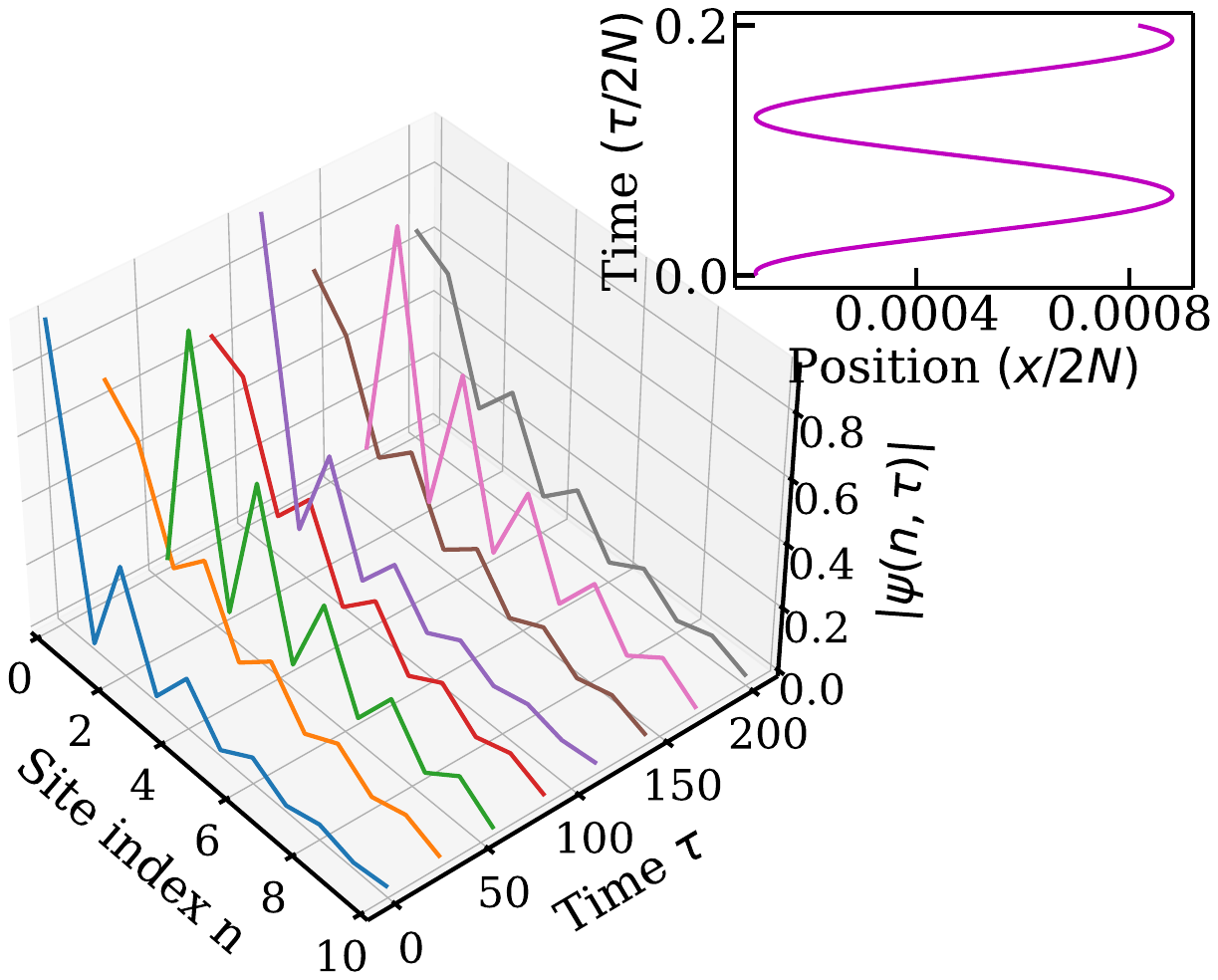}%
    \label{fig:quench and semi left}}%
  \hfill
  \subfloat[]{%
    \includegraphics[width=0.42\textwidth]{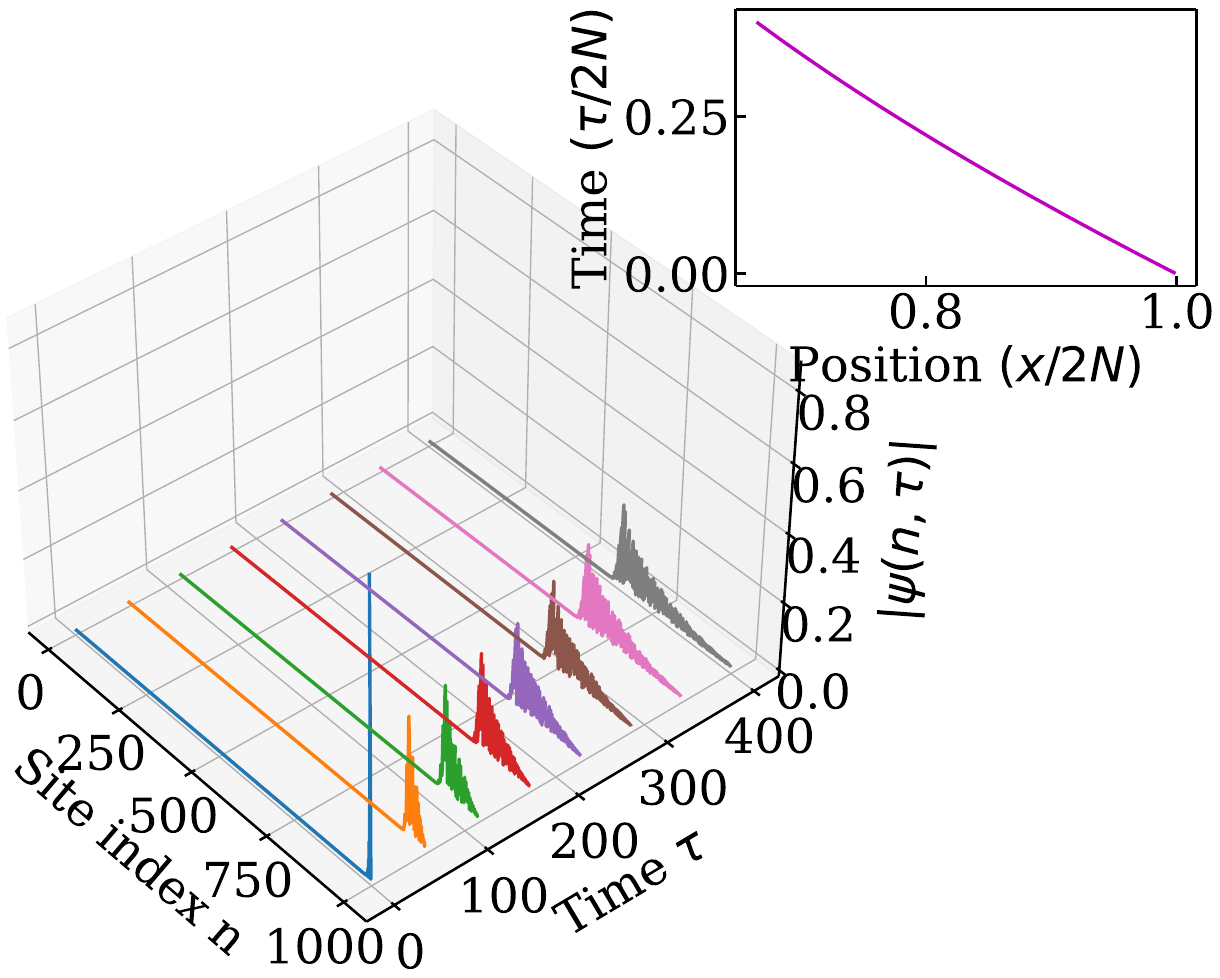}%
    \label{fig:quench and semi right}}\\[-3.8cm]
   
  \caption{Time evolution of the zero energy eigenstates for system size $2N=1000$ (500 unit cells), :  (a) left, and, (b) right edge zero energy eigenstate of a Hamiltonian with winding number 1, in another Hamiltonian with winding number zero. Inset of Fig. (a) shows the semiclassical dynamics near the origin, where the left edge state is located. Inset of Fig. (b) shows the semiclassical dynamics where the right edge state is located.}
  \label{fig:quenching and  semi}
\end{figure*}

\section{Quench dynamics: from topological to trivial phase\label{quench}}
 In this section, we study the quench dynamics across the topological transition point. For the sake of numerical simplicity and understading the mechanism better, we study the dynamics in the limit of $t_3=t_4=0$, leaving the Hamiltonian with two possible winding numbers: 0,1.  
We prepare our initial state $|\psi_i\rangle$ as one of the zero-energy eigenstates of the CST-SSH Hamiltonian for $t_2>t_1$.  Note, this is the topologically non-trivial phase, having two edge modes.  Then, we quench across the transition point in the topologically trivial phase. It implies $t_2<t_1$ for the post-quench Hamiltonian, which we identify as  $H_f$.  
The survival probability of the initial state after quench is given by \cite{PhysRevE.108.034102,PhysRevA.110.012466}, 
  \begin{equation}
      P_i(\tau)= |\langle\psi_i|e^{-iH_f\tau}|\psi_i\rangle|^2.
  \end{equation}
  It is basically a measurement of the likelihood of the zero-energy eigenstate of a topologically nontrivial Hamiltonian remaining the same after the unitary time evolution of the state in another Hamiltonian with winding number $0$. Although the topological markers produce nearly identical results for the CST-SSH and the conventional SSH models, we had identified a clear asymmetry in the edge modes of the CST-SSH model, unlike the identical edge modes of the usual SSH model. In this section, our goal is to  examine how these asymmetric edge modes evolve under our quench protocol. By contrast, in the usual SSH model the symmetric edge modes are expected to exhibit identical dynamics under the same protocol.

Figure.~\ref{fig:all_sigma} shows the variation of the survival probability with time for the usual SSH model, i.e, $\sigma=0$, and for the CST-SSH model with $\sigma=0.5$. We quench from $(t_1, t_2) =(0.5, 1)$ to $(t_1, t_2) =(1, 0.5)$. For the initial state, i.e., localized at the right edge, the time evolution of the survival probability is almost identical for the usual SSH and the CST-SSH model (see Fig.~\ref{fig:quench_0_to_1_right_sig_0.5}). The survival probability goes to zero reasonably quickly.  However, the dynamics are quite different between these two models for the left edge mode (see Fig.~\ref{fig:quench_0_to_1_left_sig_0.5}). While for the usual SSH, the dynamic is almost indistinguishable from that obtained for the right edge mode, for $\sigma >0$, the survival probability seems to oscillate between $0$ and $1$ for the CST-SSH model. We find that the oscillation period increases with increasing $\sigma$. The fact that dynamics is identical for usual SSH and not for CST-SSH is not too surprising, given that we found in Sec.~\ref{gap}  that both edge states are symmetric for $\sigma=0$, and asymmetry kicks in as soon as $\sigma >0$. 
However, it still does not explain why the survival probability of the left-edge states oscillates. Hence, we plot the absolute value of the time-evolved wavefunction in Fig.~\ref{fig:quenching and  semi}, and find that for the CST-SSH model, while under time evolution, the right edge state moves toward the other side of the lattice, while the left edge mode remains localized near its original edge. If one tracks the peak position, one finds that it moves a bit to the right and then again comes back, keeping up this to-and-fro motion. This is precisely what gets manifested in the survival probability plot. Moreover, we find the same picture even when solving the semi-classical equation of motion (see inset of Fig.~\ref{fig:quenching and  semi}).  


\captionsetup[subfigure]{position=top,skip=0pt,aboveskip=0pt,belowskip=0pt}

\begin{figure*}[!t]
  \centering
  \subfloat[]{%
    \includegraphics[width=0.40\linewidth, trim=0pt 350pt 0pt 0pt, clip]
    {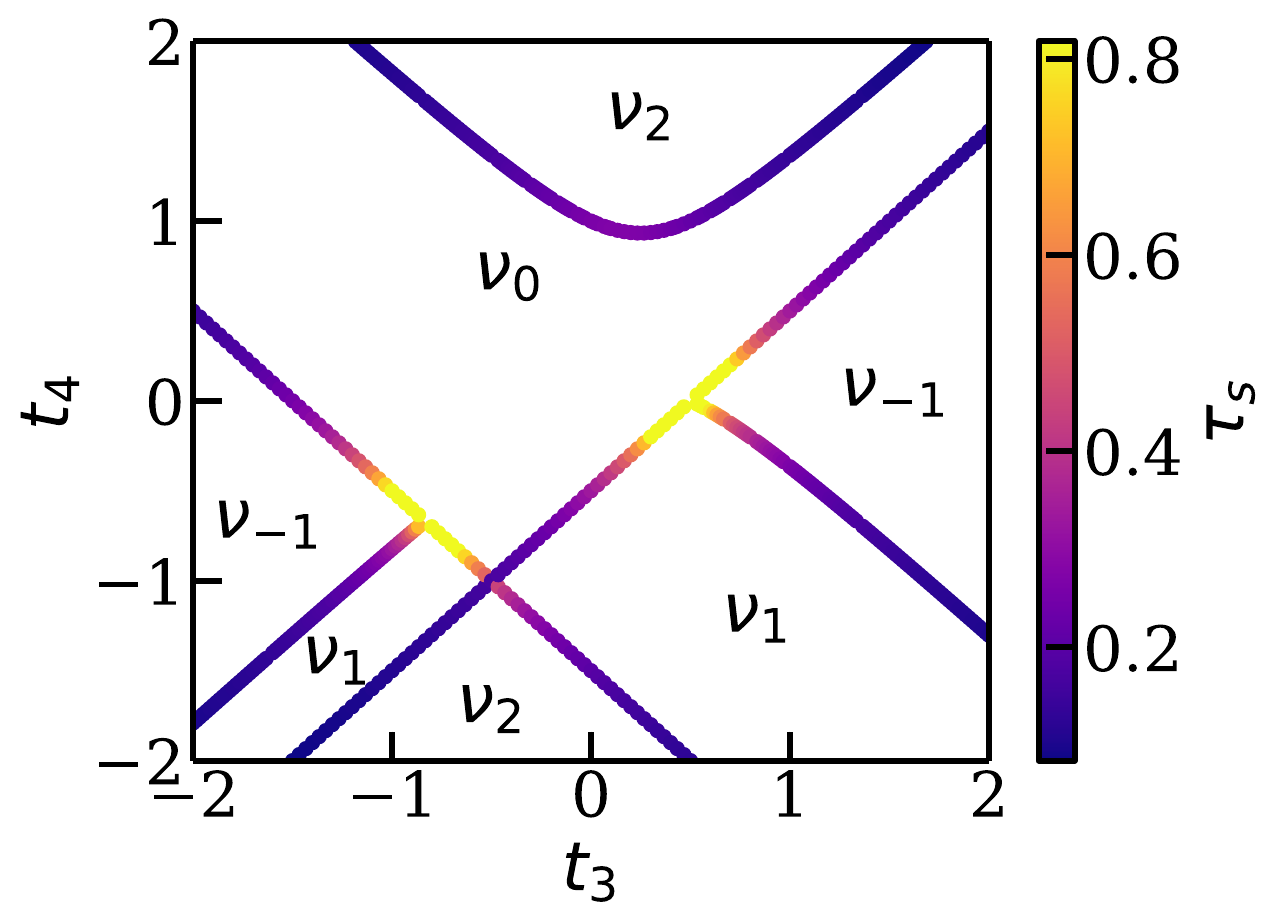}%
    \label{fig:slowdown_quantification}}%
  \hfill
  \subfloat[]{%
    \includegraphics[width=0.39\textwidth,trim=0pt 350pt 0pt 0pt, clip]{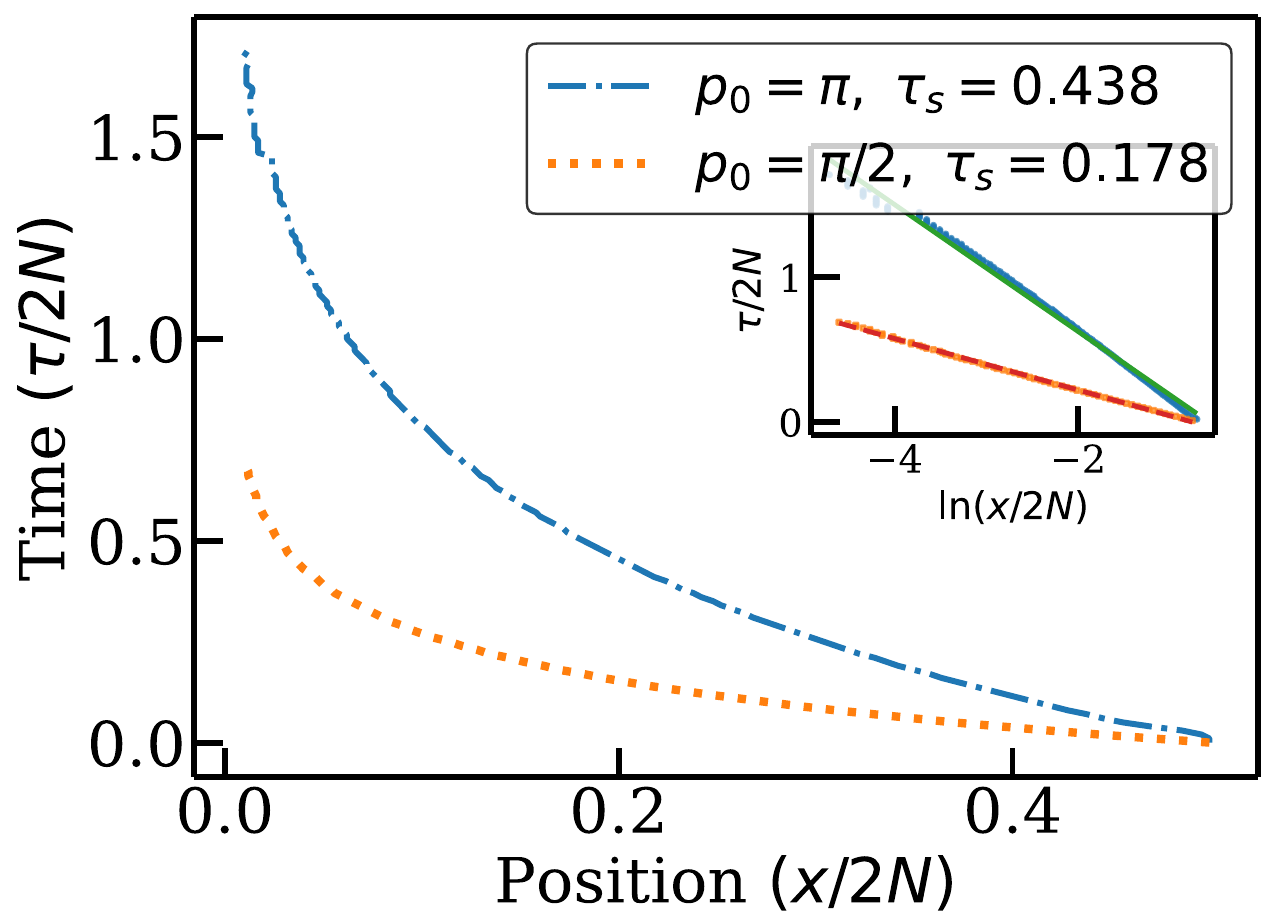}%
    \label{fig:two p0 slowdown}} 
\caption{(a)Values of slowdown time scale $\tau_s$ with $t_1=1,t_2=0.5$ and over range of $t_3,t_4$ . Here $\nu_i, (i=-1,0,1,2)$ represent regions with winding number $i$. (b) for a  multiple gap-closing point (1,0.5,-0.5,-1), two different critical slowdown parameters $\tau_s$ for two initial critical momenta, inset plot showing a Log-linear plot of the same, with straight-line fits used to extract $\tau_s$ for each $p_0$, For system size $2N=1000$ (500 unit cells). }
\end{figure*}

\section{Quantification of the critical slowdown}
\label{slowdown}


At this point, we have established that there exists a connection between the topological
transition points in SSH models and the observation of horizon-like critical slowdown in their CST
counterparts. Unlike the standard SSH model, the extended SSH model supports topological phases with higher winding numbers. While obtaining the semiclassical trajectories of the
wave packets, we observed that at different topological phase transition points the rate
at which the wave packet approaches the origin varies across the hopping parameter
space.
Therefore, in this section we quantify the slowdown rate by introducing a characteristic
time scale, $\tau_s$, which allows us to better understand the horizon-like dynamics in
topological systems with multiple gap-closing constraints for usual extended SSH models. During the critical slowdown,
the peak position of the wave packet follows
\begin{equation}
x(\tau) = x_0 e^{-\tau/\tau_s} ,
\end{equation}
as shown in Fig.~\ref{fig:two p0 slowdown}. 
Thus, for a chosen range of initial hopping parameters, we can analyze how fast or slow
the wave packet moves toward the origin.
Figure~\ref{fig:slowdown_quantification} shows the different winding numbers and the
transition lines separating the topological phases. The heatmap indicates how slowly
the wave packet moves toward the origin: larger values of $\tau_s$ correspond to
slower motion. We observe that the horizon-like critical slowdown appears only along the transition lines.

In our calculations we fix $t_1=1$ and $t_2=0.5$, and compute the slowdown parameter
$\tau_s$ over a range of $t_3$ and $t_4$. We find that when condition (iii) and either
condition (i) or (ii) of Eq.~\ref{eq:constraints} are satisfied simultaneously, the wave
packet does not evolve with time and effectively becomes stationary. The reason is that satisfying two gap-closing conditions at the same Dirac
point makes the initial group velocity vanish.
In the vicinity of these stationary points, the value of $\tau_s$ becomes larger, implying
that the wave packets approach the origin more slowly compared to other transition
points where only a single gap-closing constraint is satisfied. This behavior is visible in Fig.~\ref{fig:slowdown_quantification} as the bright yellow regions near the
junctions. We also observe another interesting phenomenon. At multiple gap-closing points of the usual extended SSH model
where both conditions (i) and (ii) of Eq.~\ref{eq:constraints} are satisfied, two different
critical slowdowns emerge for two different initial momenta $p_0$, resulting in two
distinct slowdown parameters $\tau_s$ (see Fig.~\ref{fig:two p0 slowdown}). Such
features arise only in topological systems with multiple gap-closing points, such as the
extended CST-SSH model. Note that we would like to clarify that whenever we refer to gap-closing points in this section, we mean the gap-closing points of the usual extended SSH model at the topological transition points, and not those of the CST-SSH model. This distinction is important because the CST-SSH model remains gapless for $\sigma \geq 1$, as discussed previously. Finally, this provides an additional degree of freedom to control the critical slowdown
in experimental setups designed to mimic black hole horizon-like wavepacket dynamics. Such flexibility
is absent in the simpler CST-SSH model.\\


\section{Summary and Discussion\label{summary}}

In this paper, we investigate the topological properties of both the extended and standard version of CST-SSH model. We find that, for both model, the horizon-like critical slowdown manifests only at the topological phase transition points, where a critical slowdown occurs for zero-energy wave packets near the boundary. The wave-packet dynamics results are also supported by an analytical calculation of semiclassical trajectories. 
   We analyze the energy spectrum and find that when the intercell hopping is larger than the intracell hopping, a pair of zero-energy states emerges. These states are localized at the two edges of the lattice. However, unlike in the conventional SSH model, these edge states are not symmetric. We also find that, similar to the usual SSH model, the spectrum exhibits a gap whenever the intra- and inter-cell hopping amplitudes are unequal. However, this gap scales to zero with system size as $N^{-\sigma}$. Since the typical gap of a finite-size system scales as $N^{-1}$, it implies that for $\sigma < 1$  the spectrum remains gapped in the thermodynamic limit~\cite{PhysRevB.84.115135}, and gapless for $\sigma \geq 1$.  Moreover, we use various real-space topological markers, both static and dynamic, which show a clear signature of a topological phase transition for the CST-SSH models between a topologically trivial phase and a non-trivial phase for the exact parameters, as the usual SSH models. Even though the spectrum remains gapless away from the transition points for
$\sigma \geq 1$, these regions do not exhibit horizon-like critical slowdown. This shows that
the mere gaplessness of the spectrum is not sufficient for the emergence of horizon-like wavepacket dynamics. Instead, the gapless points corresponding to topological
transitions are special, as they support horizon-like critical slowdown. This observation naturally raises a fundamental question: how can one define a
topological marker that characterizes transitions between phases that remain
gapless in the CST versions of SSH-type models? Despite the introduction of power law position dependent hopping, we show that the symmetry class of the CST-SSH models remain in BDI category. Moreover, it turns out that although the
spectrum is gapless, the states in the neighborhood of the zero-energy modes
are spatially localized (see Appendix \ref{Appendix_ext_CST_SSH}). This localization is likely the reason
why the system retains its topological character. Our IPR results suggest
effectively a mobility gap emerges in the system. In other words, the
zero-energy states are surrounded by localized states that do not contribute to transport, making the system effectively insulating. Such phenomena are
typically observed in disordered topological systems~\cite{mobility_gap_2012, PhysRevB.85.195140}. Even very recently, such phenomena have been reported in disordered SSH models, where, in contrast to the conventional SSH model, the topological transition points separate two gapless phases~\cite{mondal2025topologicalphasetransitioninfinite}.
   We also study the quench dynamics from topological to trivial state and explain the oscillatory behavior of the survival probability with our semiclassical calculations. We further analyze the slowdown quantification for the extended CST-SSH, and found that, in the vicinity of the multiple gap-closing point, where two gap closing conditions for the conventional extended SSH models are satisfied for the same $k_c$ value, the slowdown rate is higher. Also, for the other multiple gap-closing points, where two gap closing conditions are satisfied for two different $k_c$ values, we observe that there are two distinct values of $p_0$ $(=\pi, \pi/2)$, having two critical slowdown with different $\tau_s$, for the same set of hopping parameters.  With the advancement of cold-atom \cite{exp1} experiments, there is potential for our predictions to be verified in the near future~\cite{exp2,exp3}, which could enrich our understanding of both black hole physics and inhomogeneous topological systems.

   \begin{acknowledgments}
 R.M. acknowledges the DST-Inspire fellowship by the
Department of Science and Technology, Government of
India, SERB start-up grant (SRG/2021/002152). 
\end{acknowledgments}

\begin{figure*}[htbp]
\subfloat[]{%
\includegraphics[width=0.35\textwidth]{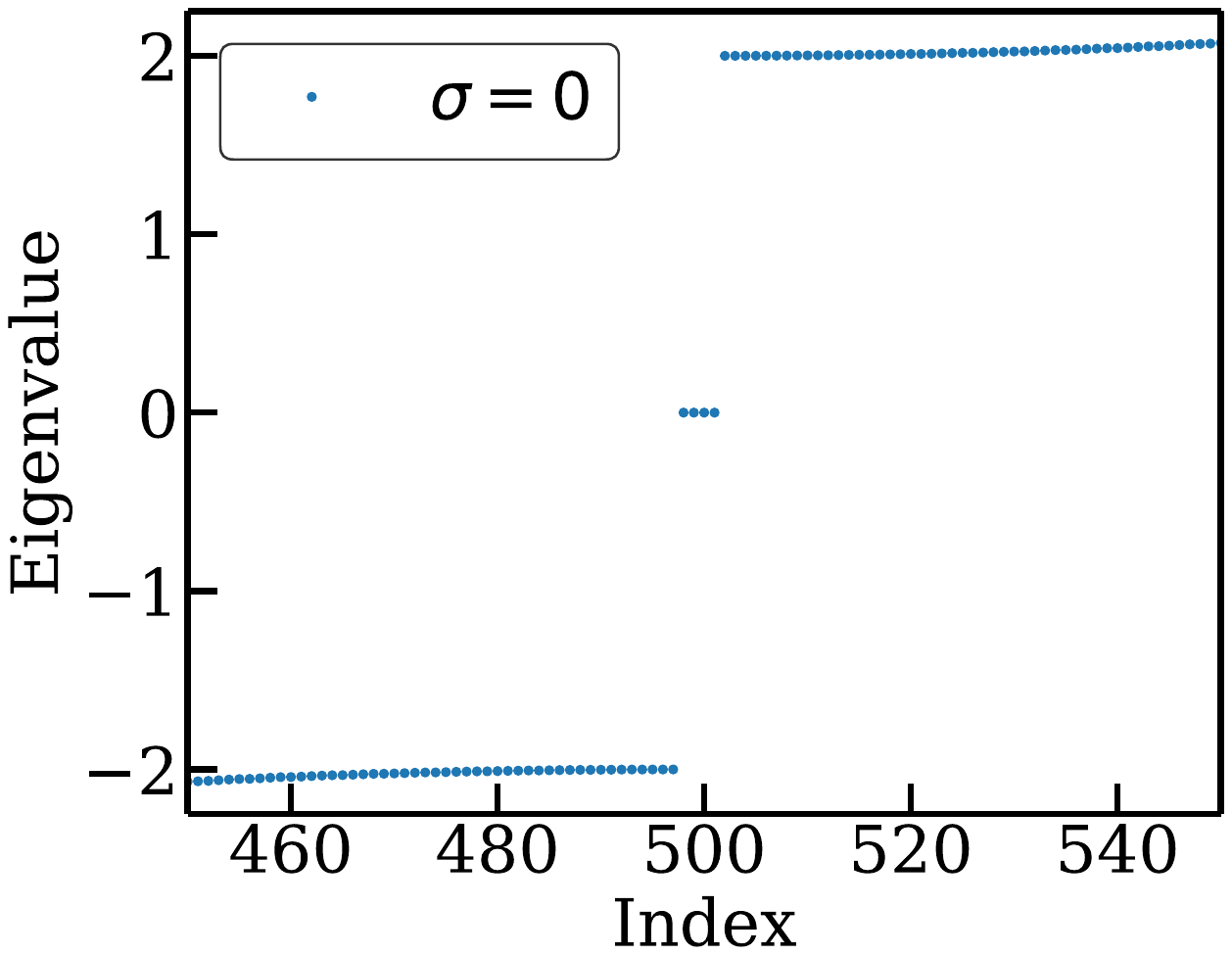}%
\label{fig:ext_0}}%
  \subfloat[]{%
\includegraphics[width=0.36\textwidth]{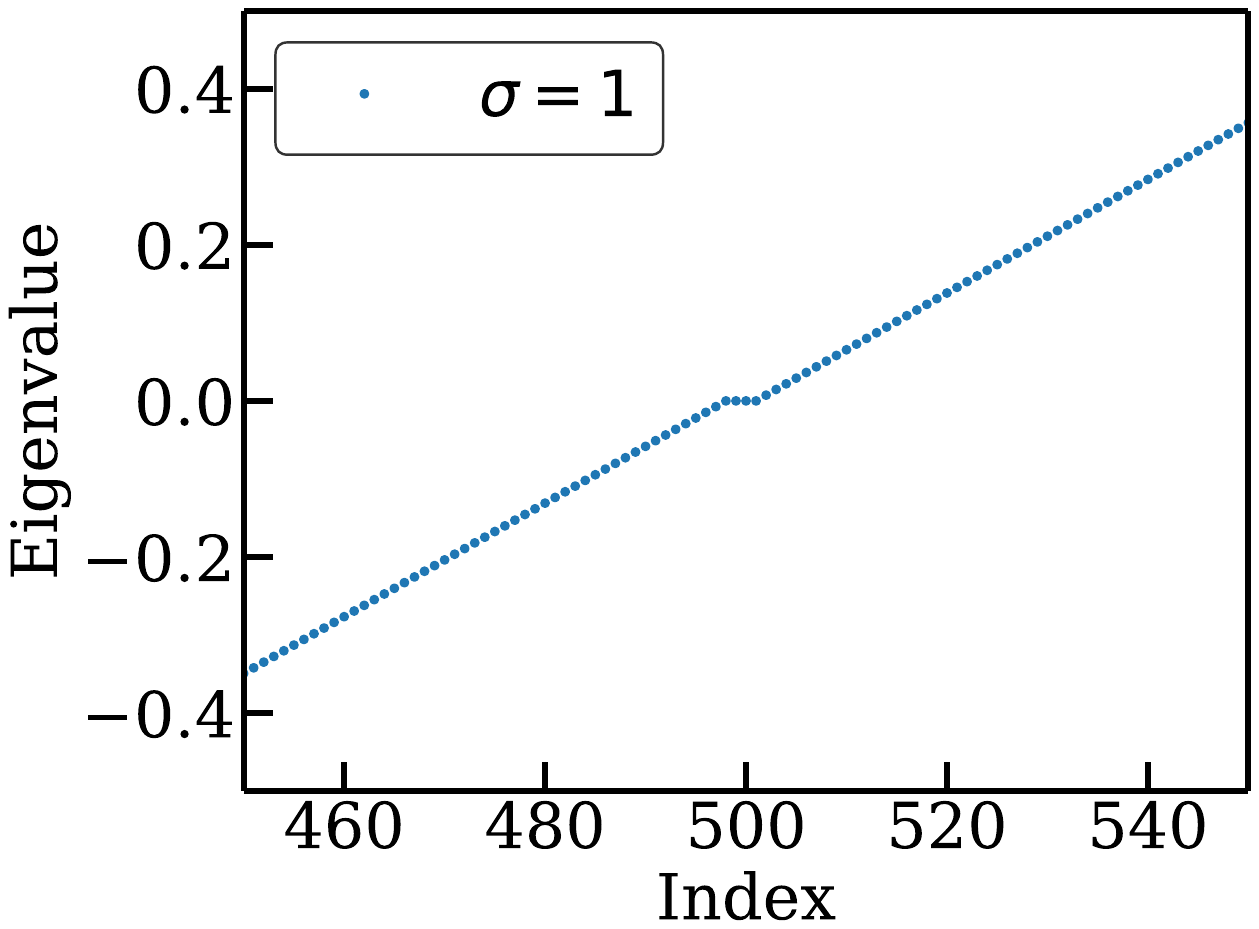}%
\label{fig:ext_1}}%
\subfloat[]{%
\includegraphics[width=0.36\textwidth]{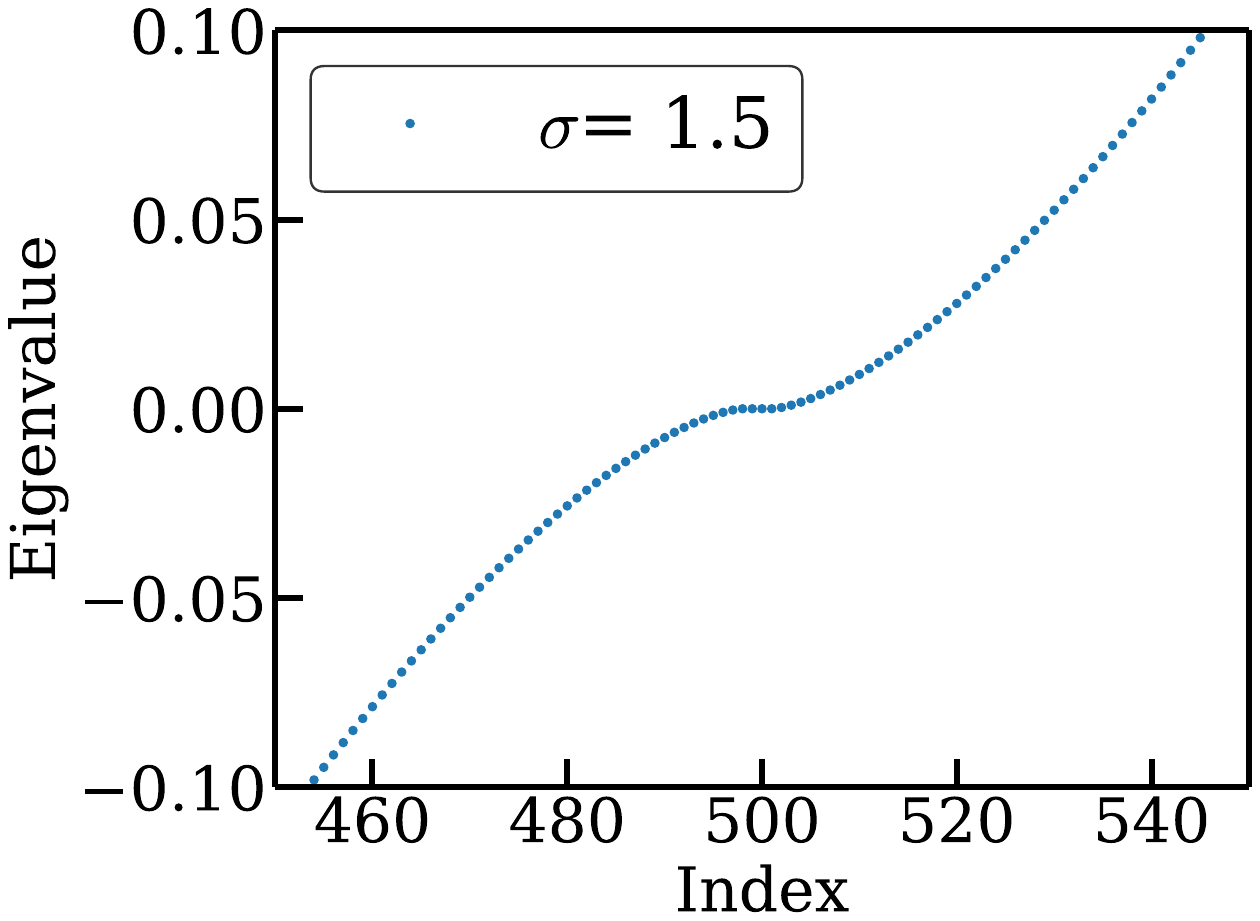}%
\label{fig:ext_1.5}}%
 \\[-3cm]

  \subfloat[]{%
    \includegraphics[width=0.35\textwidth]{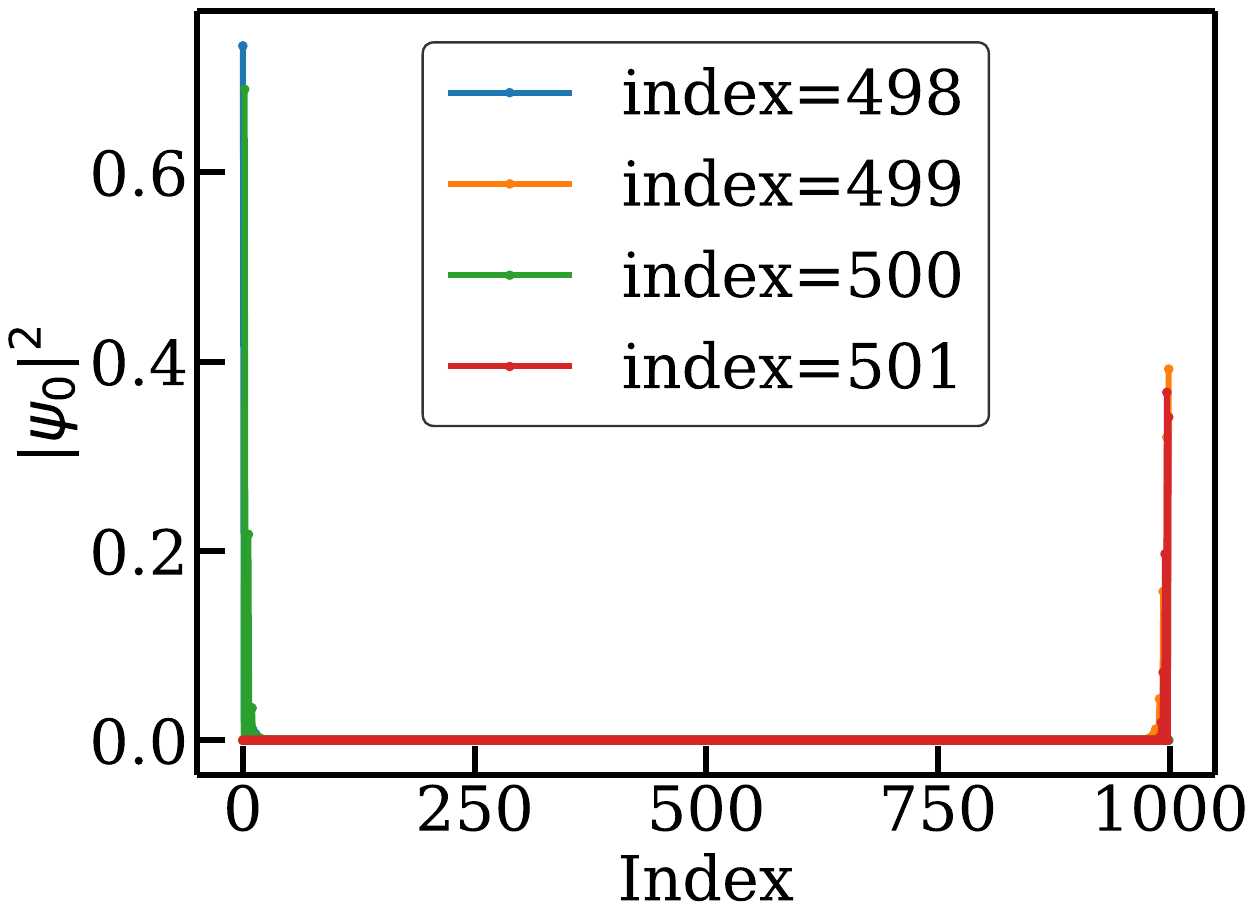}%
    \label{fig:eig_ext_0}}%
  \subfloat[]{%
    \includegraphics[width=0.35\textwidth]{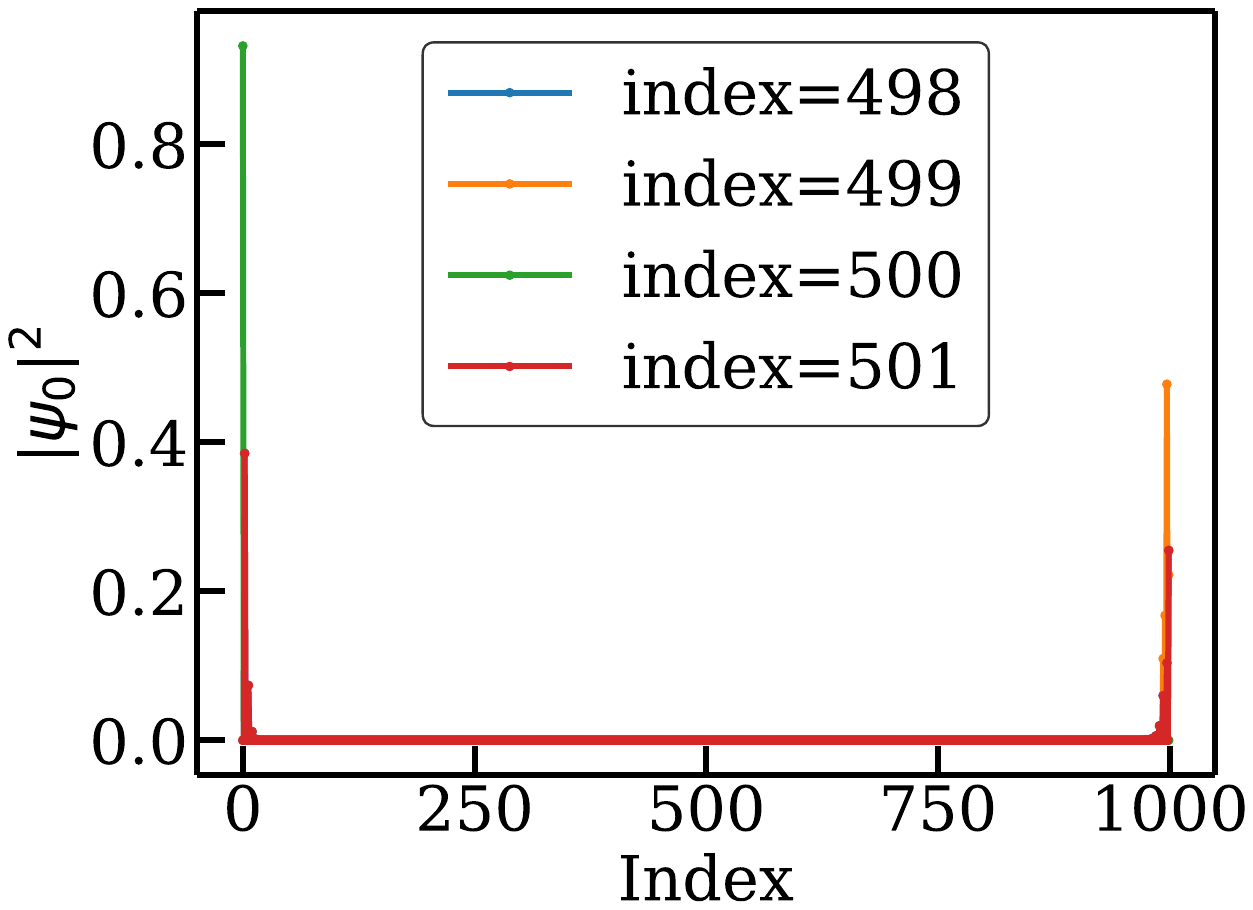}%
    \label{fig:ext_CST-SSH}}%
     \subfloat[]{%
    \includegraphics[width=0.35\textwidth]{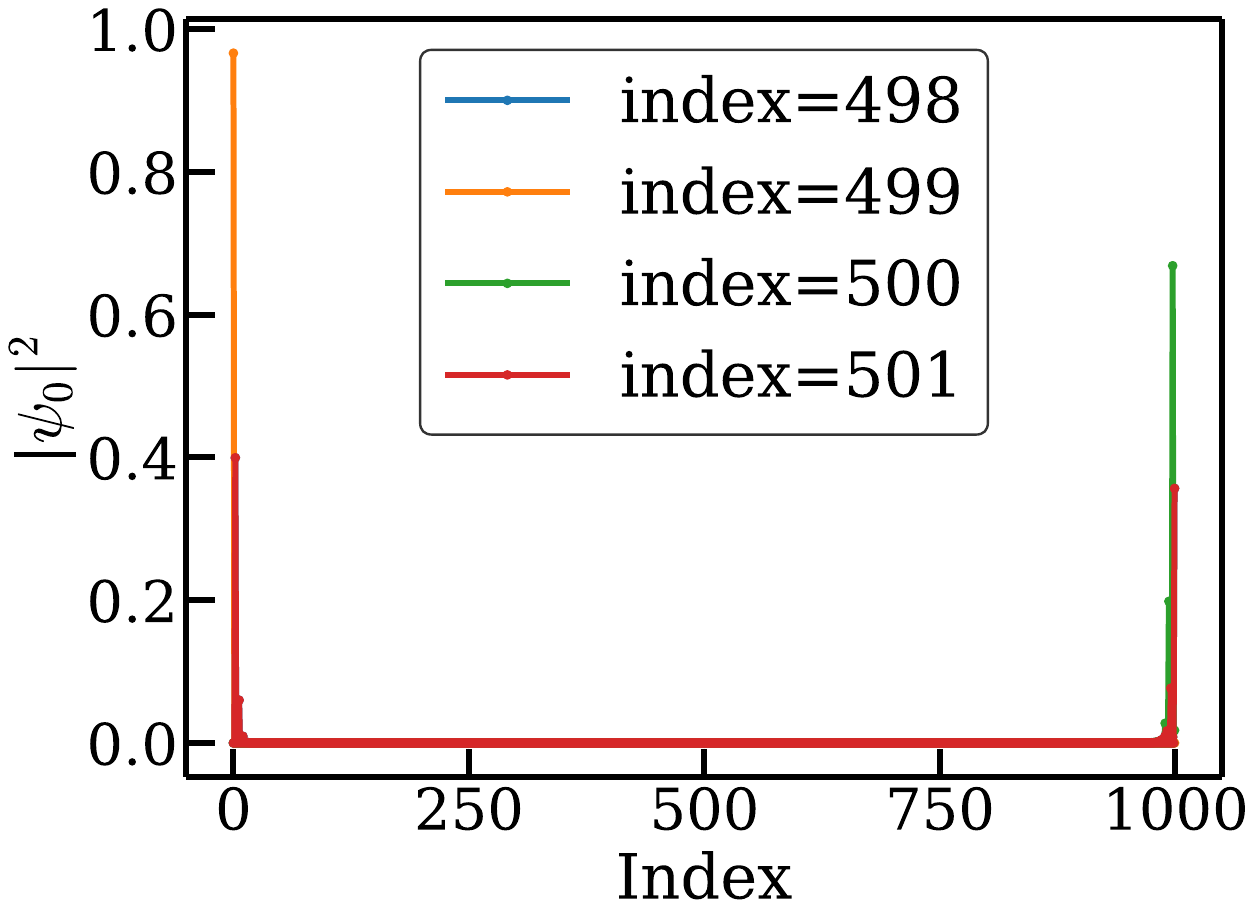}%
    \label{fig:ext_CST-SSH}}%
  \\[-3cm]
  
  \subfloat[]{%
    \includegraphics[width=0.35\textwidth]{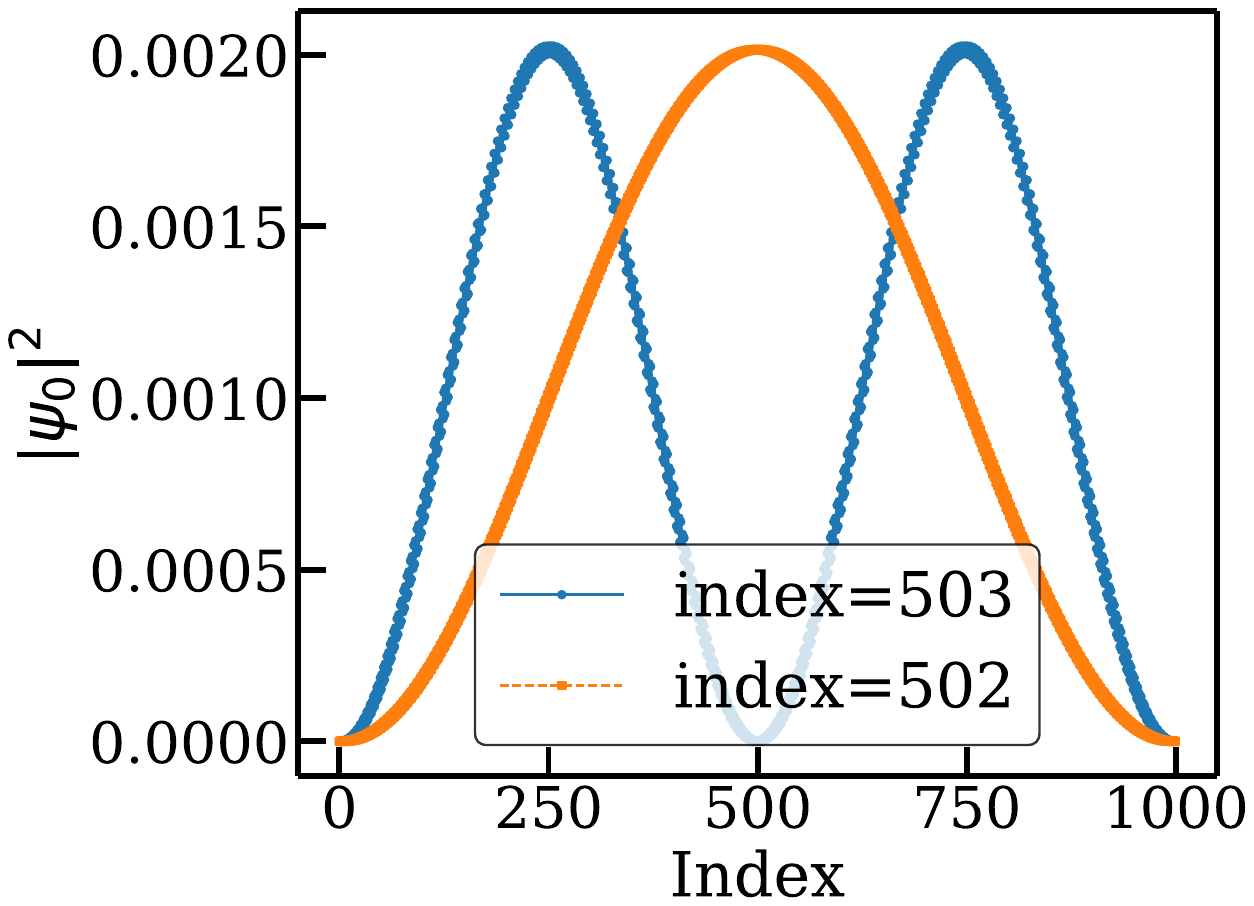}%
    \label{fig:prob_density_nonzero_eig_sig_0}}%
  \subfloat[]{%
    \includegraphics[width=0.35\textwidth]{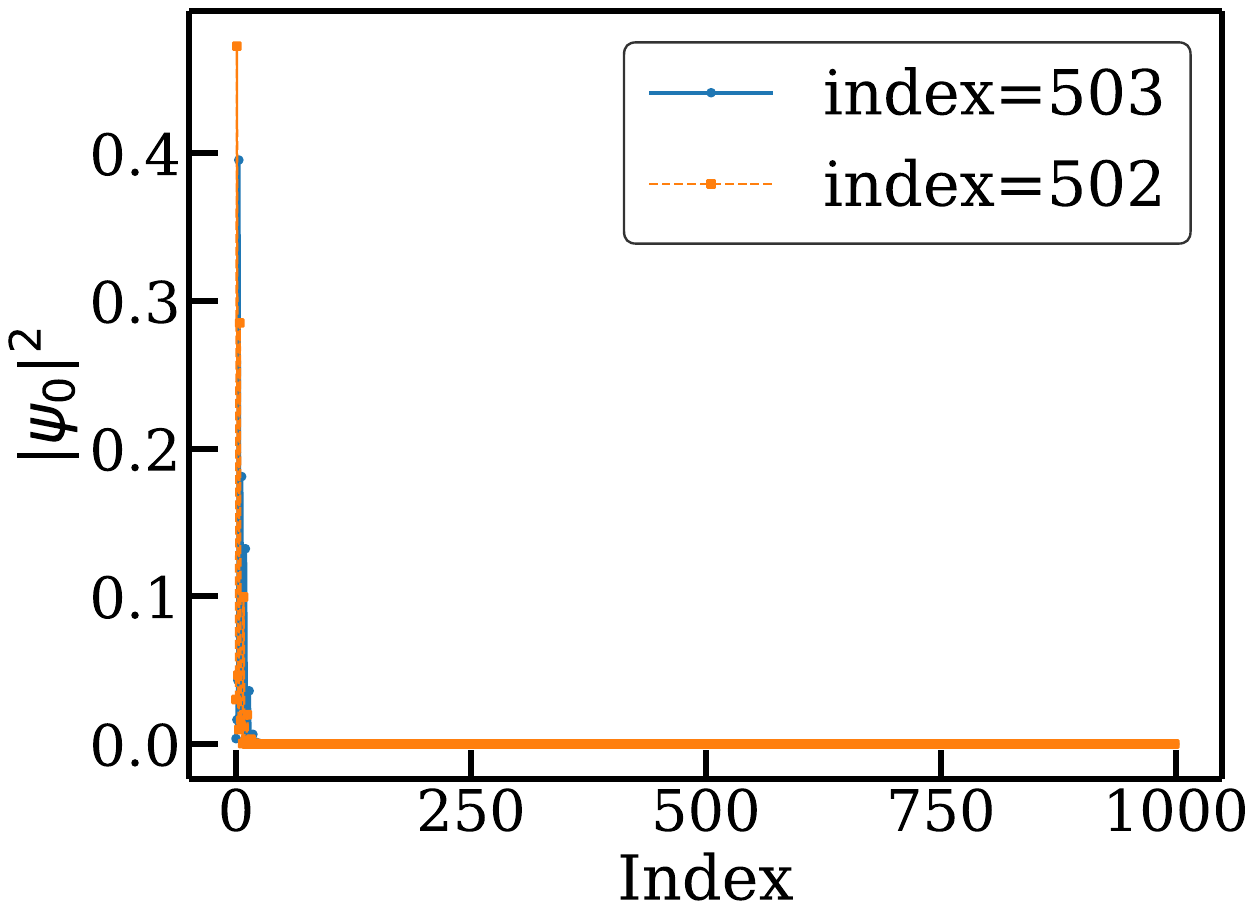}%
    \label{fig:prob_density_nonzero_eig_sig_1}}%
      \subfloat[]{%
    \includegraphics[width=0.35\textwidth]{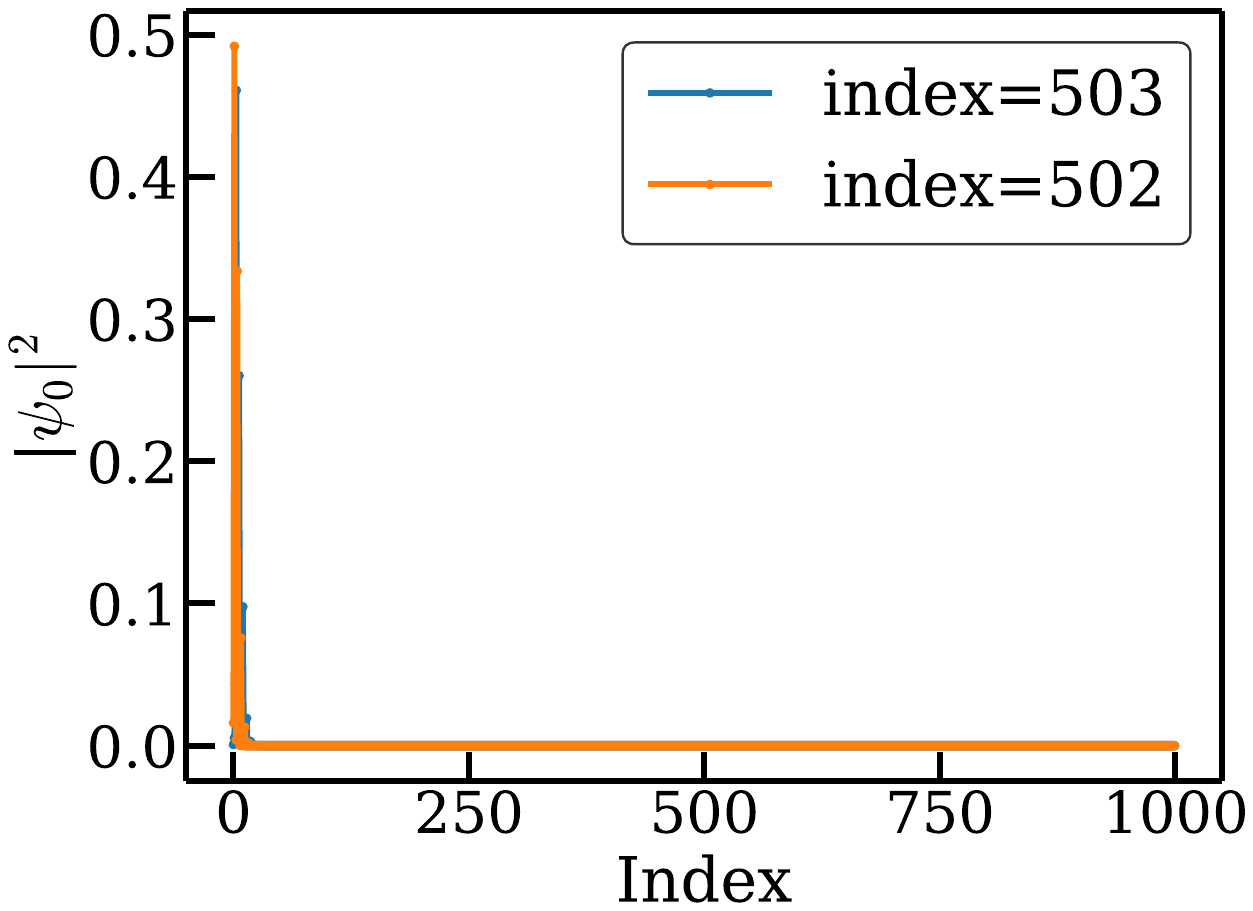}%
    \label{fig:prob_density_nonzero_eig_sig_1.5}}%
  \\[-3cm]

  \caption{Closing of the energy band gap and variation of zero energy eigen states variation with $\sigma$: for the extended CST-SSH model having system size $2N=1000$ (500 unit cells) with $(t_1,t_2,t_3,t_4)=(2,1,-1,-4)$, which have a winding number 2. Top row  (a),(b),(c): Eigenvalue spectra for $\sigma$=0, 1 and 1.5, respectively. Bottom row (d),(e),(f): Corresponding zero-energy eigenstates profiles. (g),(h), (i): probability densities for non-zero eigen energies for$\sigma=0,1 ,1 .5$, respectively}
  \label{fig:zero_energy_states_ext}
\end{figure*}

\vspace*{0.5cm}

\begin{figure*}[t]
\subfloat[]{%
\includegraphics[width=0.35\textwidth]{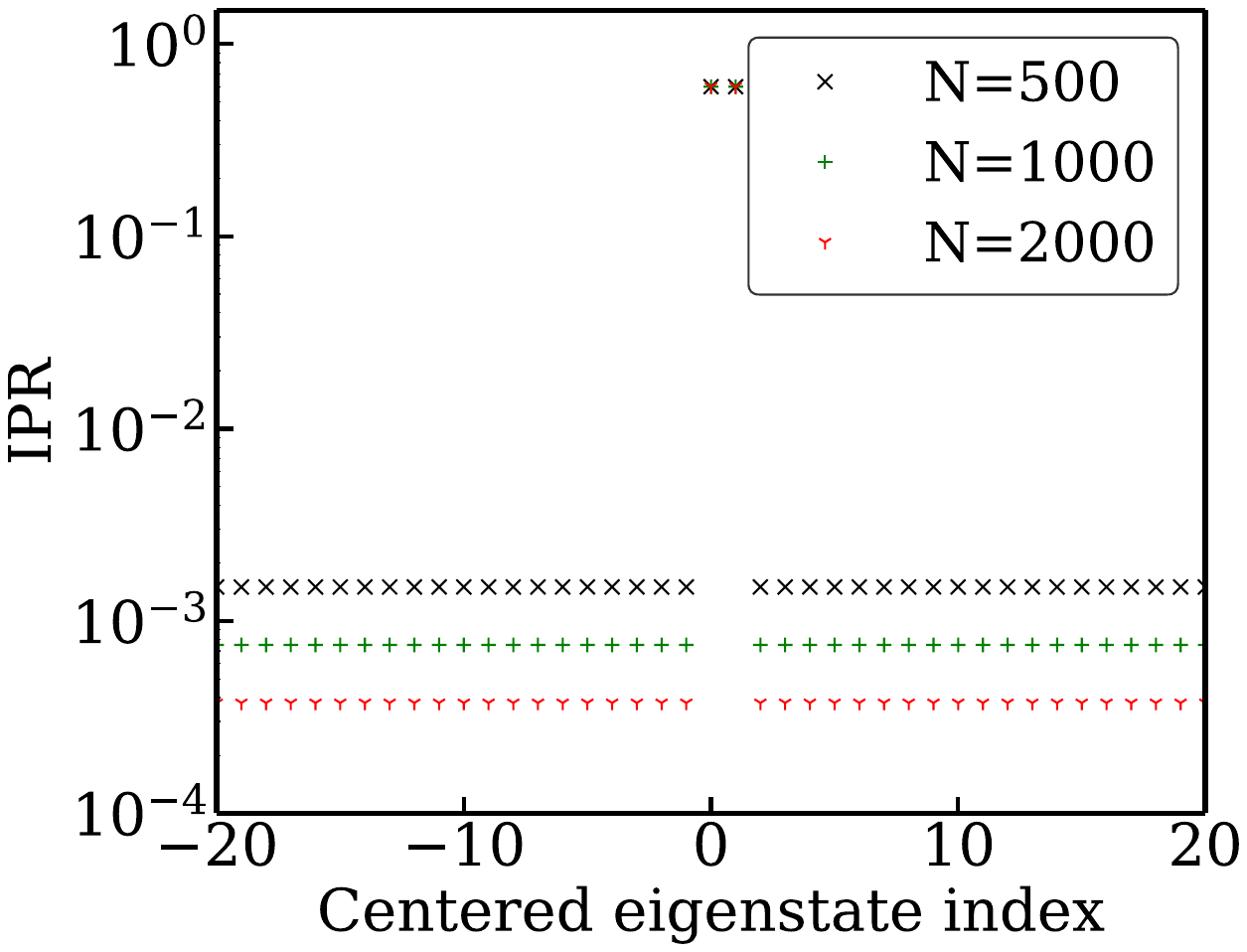}%
\label{fig:ipr_0}}%
  \subfloat[]{%
\includegraphics[width=0.35\textwidth]{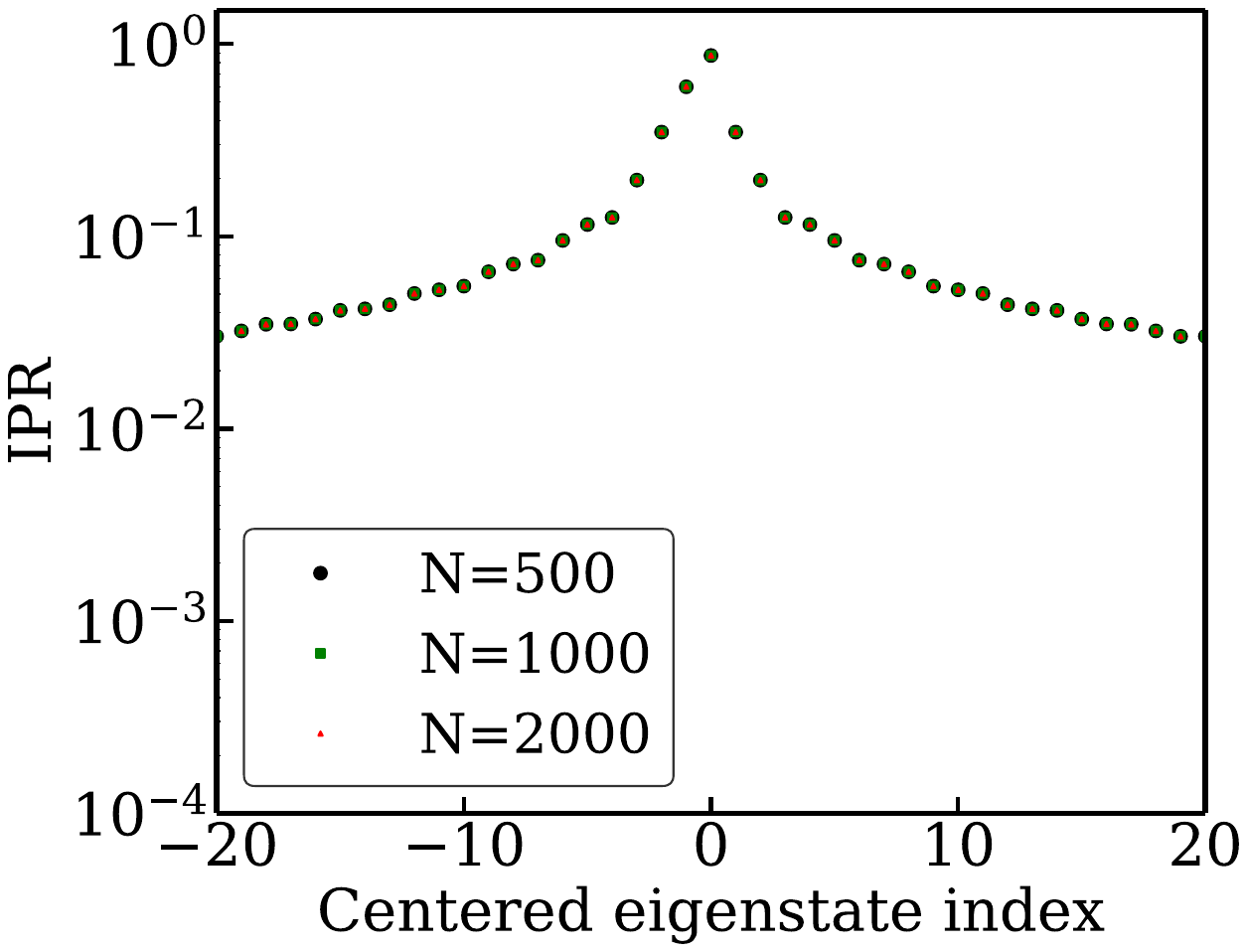}%
\label{fig:ipr_1}}%
\subfloat[]{%
\includegraphics[width=0.35\textwidth]{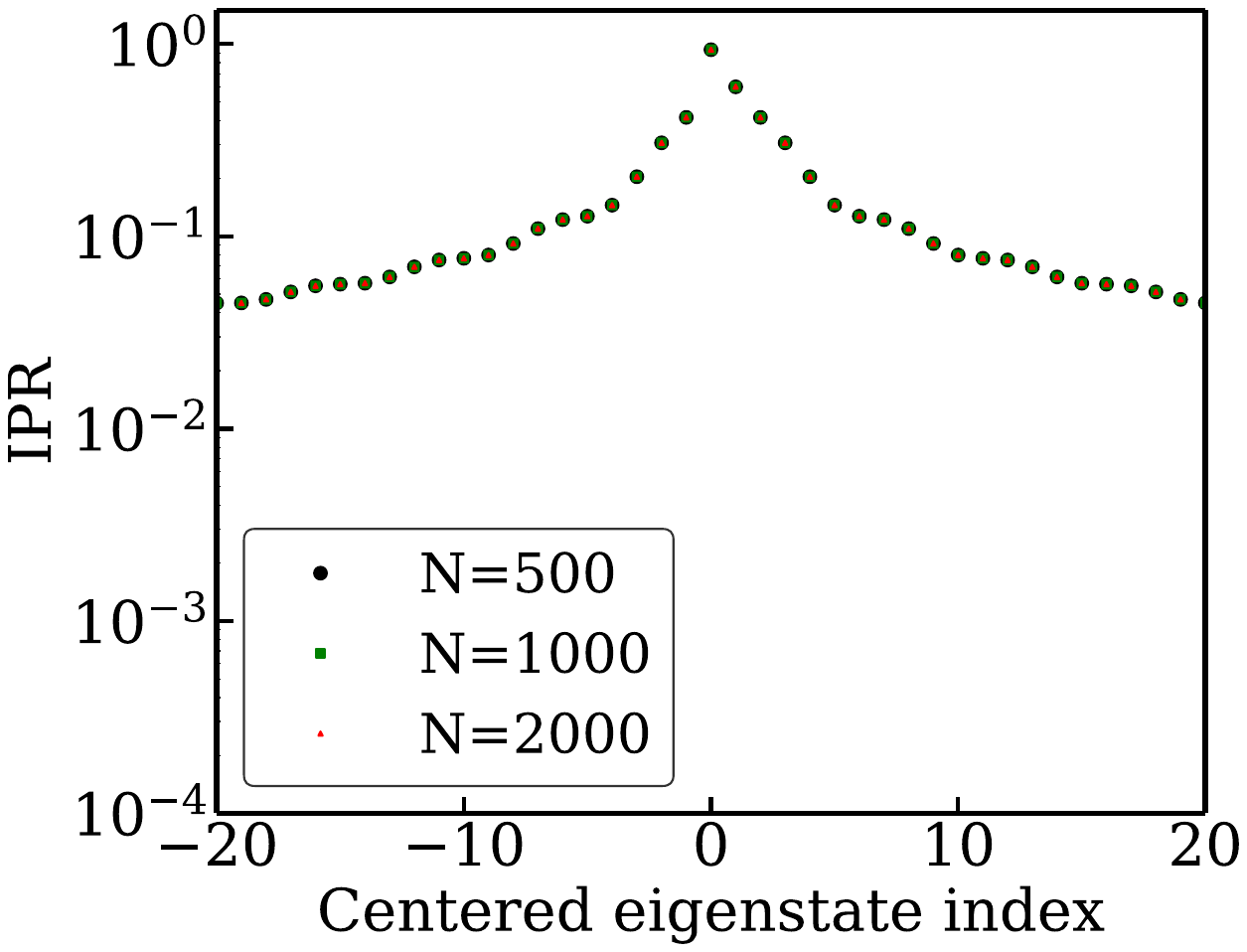}%
\label{fig:ipr_1.5}}%
 \\[-3cm]
   \caption{ IPR of the eigen states of the near zero energy neighborhood, for different system sizes with $(t_1,t_2,t_3,t_4)=(0.5,1,0,0)$, for (a) $\sigma=0$, (b) $\sigma=1$, (c) $\sigma=1.5$}
  \label{fig:ipr}
\end{figure*}

\appendix

\section{Zero energy and its neighboring states of the extended CST-SSH Model}\label{Appendix_ext_CST_SSH}

In this section, we discuss both the energy-band closing and the zero-energy states, along with their neighboring energy states, in the extended CST-SSH model.
For the Hamiltonian in Eq.~\ref{eq:ext_CST_SSH}, various winding numbers, namely $-1$, $0$, $1$, and $2$, can be realized by appropriately tuning the parameters $(t_1,t_2,t_3,t_4)$. The winding number corresponds to the number of zero-energy edge modes per boundary under open boundary conditions.

As an example, in Fig.~\ref{fig:zero_energy_states_ext}, we choose $(t_1,t_2,t_3,t_4)=(2,1,-1,-4)$, for which the winding number is equal to two. This is reflected in the presence of two zero-energy edge states at each boundary for $\sigma=0$, $\sigma=1$, and $\sigma=1.5$ (see Fig.~\ref{fig:zero_energy_states_ext} (d)-(f)). Also, as shown in Fig.~
\ref{fig:zero_energy_states_ext} (a)-(c), the energy spectra for different values of $\sigma$ are presented. For $\sigma=0$, the system exhibits a clear energy gap along with four exact zero-energy states. In contrast, for $\sigma=1$ and $1.5$, the energy gap disappears, as already discussed in the main text. Nevertheless, the four exact zero-energy states remain robust and are  visible in the spectra for both $\sigma=1$ and $1.5$. Note that with increasing $\sigma$, the energy separation between the exact zero-energy states and the neighboring states gradually decreases. As a result, the zero-energy states become less isolated from the rest of the spectrum, making them increasingly difficult to distinguish, particularly for larger values of $\sigma$.

A natural question that arises is why the model continues to support a nontrivial topology, as evidenced by the nonzero winding number reported in the main text, even after the bulk energy gap has closed. To address this issue, we examine the probability distributions of the states neighboring the exact zero-energy modes, shown in Fig.~\ref{fig:zero_energy_states_ext}(g)–(i). We find that, for $\sigma=0$, these neighboring states exhibit signatures of delocalization. In contrast, for $\sigma \geq 1$, the neighboring states remain localized. This indicates that, despite the absence of a global bulk energy gap, the system develops an effective mobility gap around zero energy, within which eigenstates remain localized. Such a mobility gap is sufficient to prevent coupling between edge and extended bulk states, thereby preserving the robustness of the zero-energy edge modes and the associated nontrivial winding number.

To quantify the degree of localization of an eigenstate $|\psi\rangle=\sum_i \psi_i |i\rangle$, we compute the inverse participation ratio (IPR), defined as \cite{mmw9-lvzd}
\begin{equation}
\mathrm{IPR}=\sum_i |\psi_i|^4,
\end{equation}
where larger values indicate stronger spatial localization. Furthermore, finite-size scaling of the IPR can be used to characterize the nature of the eigenstates. An $N$-independent IPR indicates a localized state, while an IPR that decreases with increasing system size and vanishes in the limit $N\rightarrow\infty$ corresponds to a delocalized (extended) state.
Figure~\ref{fig:ipr} presents the IPR spectrum for different system sizes and for $\sigma=0$, $1$, and $1.5$. The parameters are chosen such that the winding number is $1$, corresponding to the presence of two exact zero-energy states. To highlight the nature of these states and their nearby excitations, we restrict our attention to the spectral region in the vicinity of zero energy in Fig.~\ref{fig:ipr}.
\\For $\sigma=0$, the IPR of all states, except the two zero-energy edge modes, decreases with increasing system size, indicating that these states are extended in the thermodynamic limit. In contrast, for $\sigma=1$ and $1.5$, the states surrounding zero energy exhibit nearly system-size-independent IPR values, demonstrating that both the zero-energy edge states and a finite number of neighboring states remain localized.
Consequently, these localized states do not contribute to transport and effectively generate a mobility gap, despite the fact that the energy spectrum itself remains gapless. Such behavior is absent in the $\sigma=0$ case, where only the topological zero-energy modes are localized.

\section{Continuum Dirac limit of the extended CST-SSH model}
In this appendix, we demonstrate the correspondence between the low-energy continuum limit of the extended CST-SSH model and the Dirac equation in curved spacetime. We first consider the usual extended SSH Hamiltonian in the absence of position-dependent hopping. In momentum space, the Hamiltonian in Eq.~\ref{eq:ext_CST_SSH} can be written in the Pauli-matrix form as, 
\begin{equation}
    H(k)= d_x(k)\Sigma_1+d_y(k)\Sigma_2, 
\end{equation}
where, $d_x(k)=t_1+(t_2+t_3)\cos k+t_4\cos2k$ and, $d_y(k)=(t_2-t_3)\sin k+t_4\sin2k$ are two components of the Bloch vector. 
To derive the low-energy theory at half-filling and at the gap-closing point, we expand the Hamiltonian around one of the gap-closing points, $k_c=\pi$, and $t_1-(t_2+t_3)+t_4=0$ (see Eq.~\eqref{eq:constraints}). Introducing a small momentum deviation $q$ through $k=k_c+q$,
and retaining terms up to leading order in $q$, we obtain
\begin{equation}
    H(q)= \bigg(t_1-(t_2+t_3)+t_4+\mathcal{O}(q^2)\bigg)\Sigma_1+\bigg((t_3-t_2+2t_4)q+\mathcal{O}(q^3)\bigg)\Sigma_2
\end{equation}
Imposing the gap-closing condition
$t_1-(t_2+t_3)+t_4=0$,
at $k=\pi$ (see Eq.~\eqref{eq:constraints}), and retaining only terms linear in the momentum deviation $q$, the low-energy Hamiltonian takes the form,
\begin{equation}
    H(q)\approx (t_3-t_2+2t_4)q\Sigma_2.
\end{equation}
It is an effective massless Dirac Hamiltonian with the corresponding effective velocity $v_{\text{eff}} = (t_3-t_2+2t_4)$. We next consider the CST-SSH model with spatially varying hopping amplitudes,
$t_i(x)\approx t_i x^\sigma$.
As discussed in the main text, for sufficiently large systems the Hamiltonian can be treated locally as a position-dependent Bloch Hamiltonian $H(x,k)$. Performing a similar low-energy expansion around the gap-closing point as in the uniform case then intuitively  yields the corresponding continuum Dirac Hamiltonian,
   $H(x,q)\approx x^\sigma(t_3-t_2+2t_4)q\Sigma_2= v_{\text{eff}}(x)q\Sigma_2$. 
However, it is important to emphasize that as soon as $v_{\mathrm{eff}}(x)$ becomes position dependent, after substituting $q \rightarrow -i\partial_x$ in the position representation, the above effective Hamiltonian $H(x,q)$ is no longer Hermitian, since $
[q,v_{\mathrm{eff}}(x)] = -i\frac{dv_{\mathrm{eff}}(x)}{dx}$.
Therefore, to ensure Hermiticity, the corresponding Hermitian-conjugate term must be included. The resulting Hermitian effective Hamiltonian then takes the form
\begin{equation}
    H(x,q)\approx \frac{\Sigma_2}{2} (v_{\text{eff}}(x)q+qv_{\text{eff}}(x))
\end{equation}
Next, one obtains the corresponding equation of motion, 
\begin{equation}
     i\partial_\tau\hat\Psi=-i\Sigma_2\Big[~v_{\text{eff}}(x)\partial_x +\frac{1}{2} \frac{d v_{\text{eff}}}{dx}~\Big]\hat\Psi,
\end{equation}
which is a representation of the curved space-time massless Dirac equation as mentioned  Eq.~\ref{eq:dirac eq in cst} with $v_{\text{eff}}(x)=(t_3-t_2+2t_4)x^\sigma$, and the same metric $ds^2=-v_{\text{eff}}^2(x)d\tau^2+dx^2$ as discussed in the main text. A similar conclusion can be drawn doing an  expansion around other gap closing points. 
In the absence of the longer-range hopping terms $t_3=t_4=0$, the Hamiltonian recovers the standard CST-SSH model. The corresponding low-energy theory remains governed by a curved-space-time Dirac equation possessing the same metric form, with the only modification being a renormalized effective local velocity.

\bibliography{ref}

\end{document}